\documentstyle[12pt]{article}
\newcommand{\bec}{\begin{center}}
\newcommand{\eec}{\end{center}}
\newcommand{\lab}{\label}
\newcommand{\beq}{\begin{equation}}
\newcommand{\eeq}{\end{equation}}
\newcommand{\bea}{\begin{eqnarray}}
\newcommand{\eea}{\end{eqnarray}}
\newcommand{\nonu}{\nonumber}
\newcommand{\sma}{\sum_{n=m}^{\infty} \frac{1}{(n-m)!}}
\newcommand{\del}{\delta}
\newcommand{\dl}{\delta^{3}(l'-l)}
\newcommand{\ep}{\epsilon}
\newcommand{\om}{\omega}

\newcommand{\emm}{e_{1}. \, . e_{m}}
\newcommand{\en}{e_{1}. \, . e_{n}}

\newcommand{\dZp}{\frac{\delta^{m}}
{\delta u_{m',m}. \, .\delta u_{1',1}} Z_{P'P}}

\newcommand{\nin}{\noindent}
\newcommand{\qu}{Q_{0}}

\newcommand{\proinm}{\rho_{P}^{in}({\bf k}_{1},. \, .,{\bf k}_{m})}

\newcommand{\proexm}{\rho_{P}^{ex}({\bf k}_{1},. \, .,{\bf k}_{m})}
\newcommand{\proexn}{\rho_{P}^{ex}({\bf k}_{1},. \, .,{\bf k}_{n})}

\newcommand{\roinm}{d^{in}({\bf k}_{1}',. \, .,{\bf k}_{m}';
{\bf k}_{1},. \, .,{\bf k}_{m})}

\newcommand{\roexnm}{d^{ex}({\bf k}_{1}',. \, .,{\bf k}_{m}',
{\bf k}_{m+1},. \, .,{\bf k}_{n};{\bf k}_{1},. \, .,{\bf k}_{n})}
\newcommand{\roinmp}{d_{P}^{in}({\bf k}_{1}',. \, .,{\bf k}_{m}';
{\bf k}_{1},. \, .,{\bf k}_{m})}
\newcommand{\roexmp}{d_{P}^{ex}({\bf k}_{1}',. \, .,{\bf k}_{m}';
{\bf k}_{1},. \, .,{\bf k}_{m})}
\newcommand{\pSm}{S_{e_{1}. \, . e_{m}}(k_{1},. \, .,k_{m})}

\newcommand{\pSn}{S_{e_{1}. \, . e_{n}}(k_{1},. \, .,k_{n})}

\begin{document}
\title{ Single particle density and density matrix from the QCD cascade
        in DLA approximation
% Multiparton density matrix from the QCD cascade in DLA approximation
        \thanks{Talk given on XXXVII Cracow School of Theoretical Physics
        " Dynamics of strong interactions", Zakopane, Poland, 97}
%
% Work supported in part by the KBN grant: 2 PO 3B 08308
% and by the European Human Capital and Mobility Program ERBCIPDCT940613}
        }
\author{B. Ziaja \thanks{ A fellow of the Foundation for the Polish Science '97.
Work supported in part by the KBN grant: 2 PO 3B 08308 and by the European
Human Capital and Mobility Program ERBCIPDCT940613}
        \\
        Institute of Physics, Jagellonian University\\
        Reymonta 4, 30-059 Cracow, Poland\\
        e-mail: $beataz@thrisc.if.uj.edu.pl$}
\date{June 1997}
\maketitle
{\abstract
The structure of the QCD gluonic cascade in configuration space is investigated.
The explicit form of the inclusive single particle density in configuration
space
transverse coordinates is derived in the double logarithmic approximation (DLA)
of QCD.
The possible simplification of the multiparton density matrix formalism
for DLA approach is found and discussed.
}
%
% \newpage
%
%                               1
\section{Introduction}
The aim of this talk is to present the recent results of the investigation
\cite{bz} of the structure of the QCD gluonic cascade in configuration space.
The interest of the configuration space structure of a hadronic source has
appeared primarily in intensity interferometry \cite{HBT}. The technique
was developed originally to estimate the dimension of distant astronomical
objects. Since that time it has seen widespread application in subatomic
physics,
in particular in analysis of elementary particle collisions \cite{gelbke}.
The standard HBT procedure involves introducing an ansatz describing
the geometry of the particle source, on the basis of a physical model. Then
it investigates its multiparticle characteristcs in momentum space. Many
models of hadron production based on some phenomenological or theoretical
constraints \cite{gelbke} have been considered; however, the question as to
what is the configuration space structure of the QCD cascade when derived
explicitly
from the fundamental theory has not been addressed so far.

Recently, several groups have analysed in great detail the multiparton
distributions
in the QCD gluonic cascades \cite{others}. The results of their investigations
show that
perturbative QCD \cite{doksh} provides a powerful framework not only for the
description of
hard quark and gluon jets but also of much softer multiparticle  phenomena.
Although not understood theoretically, the hypothesis of parton-hadron duality
\cite{lphd} provides
an apparently successful link between theoretical parton distributions and
observed
particle spectra. This prescription  was extensively tested in single particle
spectra ( and total multiplicities) and found to be in a good agreement with
the available data (see e.\ g.\ \cite{doksh1}). Recently, there have appeared
indications that it may also work for
multiparticle correlations \cite{pesch1}. These unquestionable successes invite
one
to study further consequences of the theory for processes of particle
production.
At this point we would like to notice that, if one wants to exploit fully the
quantum-mechanical aspects of the QCD cascade, it is necessary to study at
first the
multiparton {\bf density matrix}. The multiparticle distributions calculated
so far
give only diagonal terms of the density matrix and thus represent a rather
restricted
( although very important ) part of the information available from the theory.

It is perhaps important to stress that in contrast to what is usually believed,
the interest in studying the multiparticle density matrix is not purely
academic. As suggested in \cite{bk}, the density matrix allows one to obtain
the multiparticle Wigner functions and consequently gives information about the
space-time
structure of the system. This allows one to make predictions about the
shape and range of the HBT interference with clear experimental consequences.

In my previous paper \cite{ziaja} I investigated the multiparticle
density matrix (DM) of the gluonic cascade produced in $e^{+}e^{-}$
collision in the framework of double logarithmic (DLA) approximation
\cite{doksh}, \cite{dla} of QCD.
I proposed a generating functional to obtain integral equations for the
multiparticle density matrix in the quasi-diagonal limit, i.\ e.\ if the
energies and emission angles of particles are close to each other.

Here I will be presenting the technique for extracting physical information
from the density matrix approach, deriving the explicit form of
inclusive single particle density in configuration space. For the sake of
simplicity, I will be restricting myself to a discussion of its dependence
on the transverse space coordinate $x_{T}$.
In solving the problem, first I will discuss the derivation of the inclusive
single particle density matrix $d_{P}^{in}(k',k)$. I will prove that the
terms in $d_{P}^{in}(k',k)$ which do not vanish for $k=k'$
are leading in the double logaritmic perturbative expansion when
$\alpha_{S} \rightarrow 0$, momentum of quark ( antiquark ) $P$
generating the gluonic cascade is large~: \newline $P\rightarrow\infty$
and the double logarithmic contributions of the form
$\alpha_{S}\, \ln^{2} P\,=\,const$ dominate. Taking into account these
leading terms will allow me
to obtain the explicit analytic form of the density matrix, and consequently
the explicit form of the single particle density in configuration space.
Finally, I will be analysing the physical properties of inclusive single
particle density, and I hope to find them in agreement with intuitive physical
expectations.
%
%                              2
\section{Density matrix formalism}
%                              2.1
\subsection{Definition of the density matrix}
In this paper I will concentrate on the single particle density in
configuration space.
I will be discussing it in relation to the QCD-parton cascade within
the framework of double logarithmic approximation (DLA), using the density
matrix formalism presented in \cite{ziaja}.
Hence in this section I will recall the definition of the density matrix
for a particle production process, and show the relation between
the density matrix and particle density in configuration space. Our assumption
is that all the produced particles are real, i.e. they are on a mass shell.
If the production of m particles can be realized in different ways
represented by a sample of Feynman diagrams, then the exclusive
m-particle density matrix equals the product of total production amplitude
$S(k_{1},. \, .,k_{m})$ and its complex conjugate
$S^{*}(k'_{1},. \, .,k'_{m})$ as~:
%%%%%%%%%%% dm 1
\beq
d^{ex}(k'_{1},. \, .,k'_{m};k_{1},. \, .,k_{m})=
S^{*}(k'_{1},. \, .,k'_{m}) S(k_{1},. \, .,k_{m});
\lab{dm1}
\eeq
where the total amplitude $S(k_{1},. \, .,k_{m})$ is the sum of all
contributions $S_{(D)}$ from Feynman diagrams (D)
multiplied by phase space factors $(2\om_{k})^{-1/2}$ :
%%%%%%%%%%% dm 2
\beq
S(k_{1},. \, .,k_{m})= \sum_{D} S_{(D)}(k_{1},. \, .,k_{m}) \prod_{i=1}^{m}
(2\om_{k_{i}})^{-1/2}.
\lab{dm2}
\eeq
For inclusive analysis one constructs the m-particle density matrix
as a series of integrated n-particle exclusive densities
in the form~:
%%%%%%%%%% dm3
\bea
\roinm = &         \nonumber\\
\sma \int [dk]_{m+1\ldots n} & \roexnm;
\lab{dm3}
\eea
where $[dk]_{i\ldots j}\equiv d^{3}k_{i}. \, . d^3k_{j}$.
This construction scheme implies, that the diagonal elements of the density
matrix
are equal to particle densities in momentum space. There is also
an obvious relation between the density matrix and particle density in real
space. Remembering, that the space-time multiparticle amplitude
is the on-mass-shell Fourier transform of the momentum amplitude, i.e.
%%%%%%%%% dm4
\beq
S(x_{1},. \, .,x_{m})= \int [dk]_{1\ldots m}
e^{i\om_{k_{1}}t_{1} - i{\bf k_{1}x_{1}}}. \, .
e^{i\om_{k_{m}}t_{m} - i{\bf k_{m}x_{m}}}
S(k_{1},. \, .,k_{m});
\lab{dm4}
\eeq
where $\om_{k_{i}}$ denotes the energy of the ith produced particle, for
the multiparticle density
$\rho^{ex/in}(x_{1},. \, .,x_{m})$ we get the relation~:
%%%%%%%%% dm5
\bea
\rho^{ex/in}(x_{1},. \, .,x_{m})=
\frac{1}{(2\pi)^{3m}} & \int & \,[dk]_{1\ldots m}
[dk']_{1\ldots m}\,d^{ex/in}(k'_{1},. \, .,k'_{m};k_{1},. \, .,k_{m})\\
& & \nonu\\
& \times & e^{i(\om_{k_{1}}-\om_{k'_{1}}) t_{1} - i( {\bf k_{1}-k'_{1}} )
{\bf x_{1} } }. \, .
e^{i(\om_{k_{m}}-\om_{k'_{m}}) t_{m} - i( {\bf k_{m}-k'_{m}} ) {\bf x_{m} } }
\nonu;
\lab{dm5}
\eea
where the factor $\frac{1}{(2\pi)^{3m}}$ was introduced to get the proper
normalization~:
%%%%%%%%% dm6
\beq
\int [dx]_{1\ldots m} \rho^{ex/in}(x_{1},. \, .,x_{m})=
\int [dk]_{1\ldots m} \rho^{ex/in}(k_{1},. \, .,k_{m}).
\lab{dm6}
\eeq
%
%
%
%%%%%%%%%%%%%%%%%%%%%%%%%%%%%%%%%%%%%%%%%%%%%%%%%%%%%%%%%%%%%%%%%%%%%
%                              2.2
\subsection{Parton cascade in the double logarithmic approximation}
The double logarithmic approach in momentum space \cite{doksh},\cite{dla}
gives a good qualitative description of the structure of the gluonic cascade.
It accounts only for the leading DL contributions to the  multiparticle
cross section.
Although emitted soft gluons violate the energy and momentum conservation rules,
however, at high energies the approximation reproduces quite well the
selfsimilar
structure of the gluon radiation. Let us consider in the framework of DLA
the gluonic cascade generated in $e^{+}e^{-}$ collision. Multiparticle exclusive
amplitude $\pSm$, describing the production of m gluons~:
%%%%%% 4 (2)
\beq
\pSm=(-1)^{n}\, e^{-\frac {w(P)}{2}} \prod_{i=1}^{m} M_{P_{i}}(k_{i})\,
e^{-\frac {w(k_{i})}{2}};
\lab{4}
\eeq
is a product of the m emission factors~:
%
%%%%%% 5 (3)
\beq
M_{P_{i}}(k_{i})=g_{S} \frac{(e_{i} \cdot
P_{i})}{(k_{i} \cdot P_{i})} \Theta_{P_{i}}(k_{i})\,G_{P_{i}};
\lab{5}
\eeq
\begin{tabbing}
where:\\
$g_{S}=\sqrt{4\pi \alpha_{s}}$, \\
n is the number of gluons emitted of quark (antiquark), \\
$k_{i}=(\omega_{i},{\bf k}_{i})$ denotes the 4-momentum of the ith soft gluon,\\
$e_{i}\equiv e_{i}^{(j)},\,j=0,\ldots,3 $ describes its polarization, \\
$P_{i}$ is the 4-momentum of the parent of the ith gluon,\\
$G_{P_{i}}$ denotes the color factor for the given vertex of the tree diagram D,
\\
\end{tabbing}
and can be conveniently represented as a tree diagram (see Fig.\ 1). Gluon
emissions are not independent. Radiated particles have the memory of their
parton parent, and of the previous parton splitting off its parent line.
This dependence restricts the phase space of the produced gluon, and it is
included in the form of a generalized step function $\Theta_{P_{i}}(k_{i})$~:
%%%%% 6 (4)
\beq
\Theta_{P}(k):= \{k^{0} \equiv \omega <P^{0},\, \theta_{{\bf kP}}<\theta, \,
\omega \theta_{{\bf kP}}>Q_{0} \};
\lab{6}
\eeq
where $P$ is the momentum of the parent of a given parton $k$, $\theta$ denotes
the emission angle of the previous parton splitting on the $P-$line, and $Q_{0}$
is a
cut-off parameter. Virtual corrections appear as a radiation Sudakov factor
$e^{-\frac {w(P)}{2}}$, where~:
%%%%% 7 (5)
\beq
w(P)=\int d^{3}k \, \langle A_{P}^{*}(k)A_{P}(k)\rangle_{(e)}
\lab{7}
\eeq
denotes the total probability of emission of a gluon from the parent P,
averaged over physical transverse polarizations $e^{1},e^{2}$.
$A_{P}(k)$ is given by~:
%%%%% 8 (6)
\beq
A_{P}(k)=\frac{M_{P}(k)}{\sqrt{2\omega_{k}}}.
\lab{8}
\eeq
It should also be emphasized that produced gluons are
real (on-mass-shell) particles, so the energy $\omega_{k}$ of a gluon of
momentum k can be approximated as~:
%%%%% 9 (7)
\beq
\omega_{k}=\mid {\bf k}\mid .
\lab{9}
\eeq
Summing of the color factors G over the color indices gives the result~:
%%%%% 10 (8)
\beq
G_{P_{i}}G^{*}_{P_{i}}=
\left \{
 \begin{array}{c}
 \mbox{$C_{F}$ dla $P_{i}=P$} \\
 \mbox{$C_{V}$ dla $P_{i}\neq P$;}
 \end{array}
\right.
\lab{10}
\eeq
where P denotes the 4-momentum of the quark (antiquark) which initializes
the gluonic cascade.

In DLA different tree diagrams come from different non-overlapping kinematic
regions, and do not interfere.
Therefore, to calculate exclusive and inclusive multigluon
densities it is enough to sum up incoherently the squares of amplitudes
(\ref{4}). Hence one obtains the exclusive density
$\proexm$~ in the form~:
%%%%% 11 (9)
\beq
\proexm=e^{-w(P)}\,\sum_{D} \prod_{i=1}^{m}
\langle A_{P_{i}}^{*}(k_{i})A_{P_{i}}(k_{i})\rangle_{(e_{i})} \, e^{-w(k_{i})};
\lab{11}
\eeq
parametrized by the momentum P of the quark
(antiquark) which initializes the cascade. Multigluon inclusive density
$\proinm$ follows from (\ref{dm3}) as~:
%%%%% 12 (10)
\beq
\proinm=\sma \int \,[dk]_{m+1 \ldots n}\,\,\proexn.
\lab{12}
\eeq
Introducing the method of the generating functional (GF) (see \cite{doksh}
and references therein) allows us to perform the summation over diagrams
in (\ref{11}) and (\ref{12}) in a very convenient way. While constructing
the generating functional $Z_{P}[u]$, one applies explicitly the selfsimilarity
property of the gluonic cascade. As a final result one obtains the recursive
{\sl master equation} in the form~:
%%%%% 13 (11)
\beq
Z_{P}[u]=e^{-w(P)}\, exp(\int d^{3}k\,
\langle A_{P}^{*}(k)A_{P}(k)\rangle_{(e)}\, u(k) Z_{k}[u]).
\lab{13}
\eeq
It can be proved \cite{doksh} that equation (\ref{13}) reproduces contributions
of all tree diagrams D, and allows us to express multigluon densities
$\proinm$ and $\proexm$ as~:
%%%%% 14 (12)
\beq
\proexm=\frac{\delta^{m}}{\delta u_{1}\ldots\delta u_{m}} Z_{P}\mid
_{\{u=0\}},
\lab{14}
\eeq
%%%%% 15 (13)
\beq
\proinm=\frac{\delta^{m}}{\delta u_{1}\ldots\delta u_{m}} Z_{P}\mid_{\{u=1\}};
\lab{15}
\eeq
where $u_{i}$ denotes the probing function $u(k_{i})$ and
the functional derivative $ \frac{\del}{\del u_{i}} $ denotes
$\frac{\del}{\del u(k_{i})} $ respectively.

We would like to emphasize the simplicity of the GF approach.
Especially for inclusive densities the method allows us to skip the complicated
summation procedure, and to express the required distribution
in a simple, compact form (for details see  \cite{wos}).
%                           2.3
\subsection{Density matrix in the DLA formalism}
The DLA formalism in momentum space gives a good description of the
structure of the gluonic cascade. The GF scheme suggests clearly
how to construct multigluon densities, and allows us to investigate their
properties in a simple way. Since the method works so well for
multiparticle distributions, one can expect to apply it successfully
for other multiparticle observables.
Below we briefly discuss the calculation of the multiparticle
density matrix in the framework of double logarithmic approximation,
recalling our main results from Ref.\ \cite{ziaja}.

First we derive the general expression for the exclusive and inclusive density
matrices $\roexmp$ and $\roinmp$. The task looks quite complicated because
in this case the interference between different diagrams in (\ref{dm1}),
generally does not vanish. Let us define two functionals, $Z_{P}[u]$ and
$Z_{P'}^{*}[w]$, which generate the sum of all tree amplitudes and
the sum of their complex conjugates respectively~:
%%%%%% 16
\bea
Z_{P}[u] & = & e^{-w(P)/2}\, exp(\int d^{3}k A_{P}(k) u(k) Z_{k}[u]),\nonumber\\
Z_{P'}^{*}[s] & = & e^{-w(P')/2}\, exp(\int d^{3}k A_{P'}^{*}(k) s(k)
Z_{k}^{*}[s]).
\lab{16}
\eea
\noindent
The multigluon density matrices can be then expressed as (see Appendix A)~:
%%%%%% 17
\bea
&  & \roexmp=\nonu\\
&  & \nonu\\
&  & = \frac{\delta^{m}}{\delta s_{1'}\ldots\delta s_{m'}}
          \frac{\delta^{m}}{\delta u_{1}\ldots\delta u_{m}}
          Z_{P}[u] Z_{P'}^{*}[s] \mid_{\{u=s=0\}, P=P'}\,\, ,
\lab{17ex}\\
&  & \nonu\\
&  & \nonu\\
&  & \roinmp=\nonu\\
&  & \nonu\\
&  &  = \frac{\delta^{m}}{\delta s_{1'}\ldots\delta s_{m'}}
          \frac{\delta^{m}}{\delta u_{1}\ldots\delta u_{m}}
          Z_{P}[u] Z_{P'}^{*}[s] \mid_{\{u=\frac{\del}{\del s}, s=0\}, P=P'}.
\lab{17}
\eea
Eq.\ (\ref{17}) is, in fact, a complicated integral equation.
The difficulty of the diagram summation does not disappear there.
It is only hidden in the compact form of (\ref{17}). However, we remember
that in the above formula we have taken into account interference
between {\sl all} different graphs D and D'. And detailed analysis gives the
result that in DLA not all the diagrams mix up: one can distinguish some
interference classes. However, at the general level we did not succeed
in formulation of such a GF which would take this fact into account.

Nevertheless, we do not need the most general form of
the density matrix. We are interested in its behaviour when the differences
of momenta \newline
$ \mid~k_{1}-k_{1}'~\mid ,\ldots, \mid~k_{m}-k_{m}'~\mid  $ are small, since
we expect that large momentum differences will not contribute to Fourier
transforms \cite{morse}. It can be shown that within this limit interferences
between different diagrams vanish, and one
sums up only "squared" contributions from identical graphs. In fact, from
the analysis of the diagrams contributing to single particle density
matrix $d_{P}^{ex}(k_{1}',k_{1})$, it has been possible to prove
(see Appendix B) that
interference between different diagrams appears only if either energies
or emission angles of produced particles are strongly ordered~:
$ \om_{1}\gg \om_{1'}$ ($ \om_{1}\ll \om_{1'}$) or
$ \theta_{1P}\gg\theta_{1'P}$ ($ \theta_{1P}\gg\theta_{1'P}$).
This statement can be generalized for any m-particle density matrix.
If we have m particles, and (m-1) ones among them are "close" to each other,
e.g.
$ k_{1}\cong k_{1}',\ldots,k_{m-1}\cong k_{m-1}'$, then the interference
of the different diagrams will take  place only if either the energies or
the angles of $ k_{m}$ and $k_{m}'$ are strongly ordered.

Hence in our approximation we exclude mixing up different
diagrams, and sum up only the squared contributions from identical ones.
The exclusive and inclusive density matrices then take the simpler form~:

\noindent
%%%%%%% 18
\bea
&  &\roexmp=\nonu\\
&  &\nonu\\
&  &\, =\sum_{D} \prod_{i=1}^{m}(4\om_{ k_{i}'}\om_{k_{i}})^{-1/2}
\,\langle S_{\emm}(k_{1}',.\, .,k_{m}')\pSm\rangle_{(\emm)},\\
\lab{18}
%%%%%%%
%%%%%%% 19
&  &\nonu\\
&  &\nonu\\
&  & \roinmp=\nonu\\
&  & \nonu\\
&  & =\sma \, \sum_{D} \int [dk]_{m+1\ldots n} \prod_{i=1}^{m}(4\om_{k_{i}'}
\om_{k_{i}})^{-1/2}
\prod_{j=m+1}^{n} (4\om_{k_{j}}\om_{k_{j}})^{-1/2}\nonu\\
&  & \nonu\\
&  & \,\,\,\,\times\, \langle S_{\en}(k_{1}',\ldots,k_{m}',k_{m+1},
\ldots,k_{n})\pSn\rangle_{(\en)}.
\lab{19}
\eea

\noindent
The summation in (\ref{18}), (\ref{19}) over D can be easily performed
using the generating functional which reproduces only contributions
of identical tree diagrams. In \cite{ziaja} I proposed
a master equation for such a generating functional (GF) $Z_{P'P}[u(k',k)]$.
It accounts only for the contributions of "diagonal" diagrams
(for proof see Appendix C), and generates the exclusive and inclusive
density matrix as~:
%%%%%%%%% dla1
\bea
 &  & \roexmp =\dZp\mid_{\{u=0\},P=P'},
\lab{dla1ex}\\
 &  & \roinmp= \dZp\mid_{\{u=\dl\}, P=P'}.
\lab{dla1}
\eea
The explicit form of its master equation, and the other formulae needed for
further calculations are presented below. The GF reads~:
%%%%%%%% dla2
\bea
Z_{P'P}[u]\,=\,e^{-W(P',P)} \sum_{n=0}^{\infty} \frac{1}{n!}
&\int&  [dk']_{1\ldots n}\, [dk]_{1\ldots n}\,
u(k'_{1},k_{1}). \, . u(k'_{n},k_{n})\nonumber\\
& & \nonu\\
& & \times \langle A_{P'}^{*}(k'_{1})A_{P}(k_{1})\rangle_{(e_{1})}
. \, . \langle A_{P'}^{*}(k'_{n})A_{P}(k_{n})\rangle_{(e_{n})}\nonu\\
& & \nonu\\
& &  \times Z_{k'_{1}k_{1}}[u] . \, .
Z_{k'_{n}k_{n}}[u]\,{\em P}_{1',. \, .,n';1,. \, .,n};
\lab{dla2}
\eea
where the function ${\em P}_{1',. \, .,n';1,. \, .,n}$ provides
the requested parallel angular ordering  ( see Fig.\ 5)
for n particles in the form~:

%%%%%%%%%% dla3
\beq
{\em P}_{1',. \, .,n';1,. \, .,n}=
\sum_{(i_{1},. \, .,i_{n}) \in Perm(1,. \, .,n)}
\Theta( \theta_{k'_{i_{1}}P'}>. \, .>\theta_{k'_{i_{n}}P'})\,\,
\Theta( \theta_{k_{i_{1}}P}>. \, .>\theta_{k_{i_{n}}P});
\lab{dla3}
\eeq
\noindent
the product of the single particle amplitudes
$\langle~A_{P'}^{*}(k') A_{P}(k)\rangle_{(e)}$
averaged over gluon polarizations (see Appendix D) reads~:
%%%%%%%%% dla4
\bea
 &  & \langle A_{P'}^{*}(k')A_{P}(k)\rangle_{(e)} \equiv A_{P'P}(k',k) =
 \nonumber\\
 &  & \nonu\\
 &  & =\frac{4g_{S}^{2}}{2\pi}\, G_{P'}^{*}G_{P}\,
 \frac{1}{\sqrt{4\om^{\prime 3}\om^{3}}}
\frac{1}{\theta_{Pk}\theta_{P'k'}} \Theta_{P'}(k')\Theta_{P}(k);
\lab{dla4}
\eea
\noindent
and the radiation factor is given by~:
%%%%%%%% dla5
\beq
W(P',P)=\frac{w(P')+w(P)}{2}.
\lab{dla5}
\eeq
\noindent
The other notations are the same as in section 2.\ 2.\

Generating functional (\ref{dla2}) describes the sequence of
"emissions"~: \newline $\langle A_{P'}^{*}(k')A_{P}(k)\rangle_{(e)}$ of
two particles $k$, $k'$ from two parents $P$, $P'$. If the profile
function $u(l',l)$ is equal  to $0$ and $\dl$ respectively,
then the functional $Z_{P'P}$ takes the form~:
%%%%%% dla10
\bea
Z_{P'P}[u=0]=  &  e^{-W(P',P)},\nonumber\\
Z_{PP}[u=\dl]=  &  1.
\lab{dla10}
\eea
For the profile function of the form $u(l',l)=v(l)\, \dl$ and $P=P'$
our $Z_{P'P}[u]$ reduces to functional (\ref{13}), as expected~:
%%%%%% dla101
\beq
Z_{PP}[u(l',l)=v(l)\,\dl]\,= \,Z_{P}[v(l)].
\lab{dla101}
\eeq
%                              2.4
%
\subsection{Single particle density matrix. Leading terms}
From relations (\ref{dla1}), (\ref{dla2}) one obtains a simple
integral equation for the inclusive single particle density matrix~:
%%%%%%%% dla11
\bea
d_{P}^{in}(k';k) & = & \int d^{3}s A_{PP}(s,s)
d_{s}^{in}(k';k)+\nonumber\\
 & + & f_{P}(k',k) A_{PP}(k',k)\,Z_{k',k}[u=\dl];
\lab{dla11}
\eea
where $f_{P}(k',k)$ equals
%%%%%%%% dla12
\bea
f_{P}(k',k) & = & e^{-g_{S}^{2}C_{F}\mid\ln^{2}
\frac{\theta_{kP} P}{ Q_{0}}-\ln^{2}
\frac{\theta_{k'P} P}
{ Q_{0}}\mid}.
\lab{dla12}
\eea
Introducing the notation~:
%%%%%%% dla13
\beq
g_{P}(k',k)\equiv f_{P}(k',k)\, A_{PP}(k',k)\,Z_{k',k}[u=\dl];
\lab{dla13}
\eeq
we may write the symbolical solution $d_{P}^{in}(k',k)$ of (\ref{dla11})
in the form~:
%
%%%%%%% dla14
\beq
d_{P}^{in}(k',k)= \sum_{n=0}^{\infty}\int d^{3}s_{1}.\, . d^{3}s_{n}\,
A_{PP}(s_{1},s_{1}).\, .A_{s_{n-1}s_{n-1}}(s_{n},s_{n}) \, \, g_{s_{n}}(k',k).
\lab{dla14}
\eeq
We present Eq.\ (\ref{dla14}) as a final result obtained in Ref.\ \cite{ziaja}.
The exact solution of $d_{P}^{in}(k',k)$ and its detailed properties will be
discussed below.
We may replace $A_{PP}(s,s)$ in (\ref{dla14}) with its explicit form
taken from Eq.\ (\ref{dla4}):
%%%%%%ie2
\bea
d_{P}^{in}(k',k)&=&
\sum_{n=0}^{\infty} (2b)^{n}
\int \frac{ds_{1}}{s_{1}}
\int \frac{d\Omega_{s_{1}}}{2\pi\theta_{Ps1}^{2}} \Theta_{P}(s_{1}).\,.
\int \frac{ds_{n}}{s_{n}}
\int
\frac{d\Omega_{s_{n}}}{2\pi\theta_{n-1,n}^{2}}\Theta_{s_{n-1}}(s_{n})\nonu\\
& & \nonu\\
& & \hspace*{\fill}\times\,g_{s_{n}}(k',k);
\lab{ie2}
\eea
where $\int d\Omega_{s_{i}}$ denotes the angular integration over the direction
of vector ${\bf s_{i}}$, $\Theta_{i,j}\,\equiv\,\Theta_{s_{i},s_{j}}$
and $b\equiv g_{S}^{2}C_{F}$.
Factor $g_{s}(k',k)$ defined by expression (\ref{dla13}), reads~:
%%%%%%ie3
\bea
g_{S}(k',k) & = & \frac{2b}{2\pi}\,\frac{1}{\sqrt{k^{3}k^{\prime 3}}}
\frac{1}{\theta_{Sk}\theta_{Sk'}}\Theta_{S}(k')
\Theta_{S}(k)\times\nonumber\\
&& \nonu\\
&\times & e^{-b\mid\ln^{2} \frac{\theta_{kS} S}{ Q_{0}}-\ln^{2}
\frac{\theta_{k'S} S}
{ Q_{0}}\mid}\,
e^{-b'/2(\ln^{2} \frac{k\theta_{kS}}{ Q_{0}}+\ln^{2} \frac{k'\theta_{k'S} }
{ Q_{0}})}\nonu\\
&& \nonu\\
&\times & \sum_{n=0}^{\infty} (2b')^{n} \int [dk]_{1\ldots n}\,
\frac{1}{2\pi\theta_{k1}\theta_{k'1}}. \, .
\frac{1}{2\pi\theta_{kn}\theta_{k'n}}
\, \frac{1}{k_{1}^{3}}. \, .\frac{1}{k_{n}^{3}}\,\nonu\\
&& \nonu\\
& \times & \Theta( \theta_{k1}>. \, .>\theta_{kn})
\Theta( \theta_{k'1}>. \, .>\theta_{k'n})\nonumber\\
&& \nonu\\
& \times & \Theta_{k'}(1). \, .\Theta_{k'}(n)\Theta_{k}(1). \, .\Theta_{k}(n);
\lab{ie3}
\eea
where $b'\equiv g_{S}^{2}C_{V}$. Rewriting its phase space restrictions in
terms of Bessel functions (see Appendix E) one arrives at~:
%%%%%% ie4
\bea
g_{S}(k',k)& = &\frac{2b}{2\pi}\,\frac{1}{\sqrt{k^{3}k^{\prime 3}}}
\frac{1}{\theta_{Sk}\theta_{Sk'}}
\Theta_{S}(k') \Theta_{S}(k)\times\nonu\\
& &\nonu\\
& \times & e^{-b\mid\ln^{2} \frac{\theta_{kS} S}{ Q_{0}}-\ln^{2}
\frac{\theta_{k'S} S}
{ Q_{0}}\mid}\,
e^{-b'/2(\ln^{2} \frac{k\theta_{kS}}{ Q_{0}}+\ln^{2} \frac{k'\theta_{k'S} }
{ Q_{0}})}\nonu\\
& &\nonu\\
& \times & \sum_{n=0}^{\infty} (2b')^{n}\,
\prod_{i=1}^{n}\,
\left( \int_{0}^{\infty}dx_{i} x_{i} J_{0}(x_{i}\theta_{k'k})\, \right.
\int_{max(\frac{\qu}{\theta_{k,i-1(k',i-1)}})}^{min(k,k')}\,
\frac{dk_{i}}{k_{i}}\nonu\\
& & \nonu\\
& \times & \int_{\frac{\qu}{k_{i}}}^{\theta_{k,i-1}} d\theta_{ki}
J_{0}(x_{i}\theta_{ki})
\left. \int_{\frac{\qu}{k_{i}}}^{\theta_{k',i-1}} d\theta_{k'i}
J_{0}(x_{i}\theta_{k'i}))\right).
\lab{ie4}
\eea

The term which dominates in (\ref{ie3}) in the double logarithmic perturbative
expansion~: $b\,\rightarrow 0$, $P\rightarrow\infty$,
$b \ln^{2}\frac{P}{\qu}=const$, has the explicit form
(for details see Appendix F )~:
%%%% ie5
\beq
g_{S}^{(1)}(k',k)=
\frac{2b}{2\pi}\,\frac{1}{\sqrt{k^{3}k^{\prime 3}}}\,\frac{1}{\theta_{Sk'}
\theta_{Sk}}
\Theta_{S}(k')\, \Theta_{S}(k).
\lab{ie5}
\eeq
One may check (see Appendix F ) that this term iterated in (\ref{ie2}) produces
the DLA leading contribution to the density matrix $d_{P}^{in}(k',k)$.
Namely, if we introduce the notation~:
%%%%% ie6
\bea
d_{P}^{in,\,(1)}(k',k)& \equiv &
\sum_{n=0}^{\infty} (2b)^{n}
\int \frac{ds_{1}}{s_{1}}
\int \frac{d\Omega_{s_{1}}}{2\pi\theta_{P1}^{2}} \Theta_{P}(s_{1}).\,.
\int \frac{ds_{n}}{s_{n}}
\int
\frac{d\Omega_{s_{n}}}{2\pi\theta_{n-1,n}^{2}}\Theta_{s_{n-1}}(s_{n})\nonu\\
& & \nonu\\
& & \hspace*{\fill}\times\,g_{s_{n}}^{(1)}(k',k);
\lab{ie6}
\eea
then there is a relation~:
%%%%%ie7
\bea
d_{P}^{in}(k',k)\,\leq\,d_{P}^{in,\,(1)}(k',k)\lab{ie7};
\eea
and for $k \cong k'$~in quasi-diagonal limit~:
%%%%%ie8
\beq
d_{P}^{in}(k',k)\,\cong \,d_{P}^{in,\,(1)}(k',k).
\lab{ie8}
\eeq
Both (\ref{ie7}) and (\ref{ie8}) hold in the perturbative limit, as well.
Hence we may write finally that~:
%%%%%% d3
\bea
d_{P}^{in}(k',k)& \stackrel{\rm DLA}{=} &
\sum_{n=0}^{\infty} (2b)^{n}
\int \frac{ds_{1}}{s_{1}}
\int \frac{d\Omega_{s_{1}}}{2\pi\theta_{P1}^{2}} \Theta_{P}(s_{1}).\,.
\int \frac{ds_{n}}{s_{n}}
\int
\frac{d\Omega_{s_{n}}}{2\pi\theta_{n-1,n}^{2}}\Theta_{s_{n-1}}(s_{n})\nonu\\
& & \nonu\\
& & \hspace*{\fill}\times\,g_{s_{n}}^{(1)}(k',k).
\lab{d3}
\eea
The above result simplifies significantly the single particle density matrix
approach \cite{ziaja}. We apply it in section 4.\ 2.\ so as to improve
calculation of the single particle density in mixed coordinates.
%
%
%%%%%%%%%%%%%%%%%%%%%%%%%%%%%%%%%%%%%%%%%%%%%%%%%%%%%%%%%%%%%%%%%%%%%%%%%%
%                              3
\section{DLA in configuration space}
%                              3.1
\subsection{Fourier transform in transverse coordinates}
The DLA soft parton cascade starts from the initial parton of
momentum ${\bf P}$. Momenta of all particles produced from parton
$P$ refer to the ${\bf P}$ direction: they depend on the
{\bf transverse and longitudinal} momenta ${\bf k_{T}}$ and $k_{L}$
(see e.\ g.\ \cite{doksh}, \cite{mueller}) taken with respect to the {\bf P}
axis.

This dependence influences the structure of the cascade in configuration space.
For the single particle amplitude Eq.\ (\ref{dm4}) takes the form~:
%%%%% re1
\beq
S_{P}(x)= \int d^{3}k\,
e^{i\om t - i{\bf k x}}
S_{P}({\bf k})=
\int dk_{L}d^{2}k_{T}\,
e^{i\om t - i{\bf k x}}
S_{P}({\bf k})
;
\lab{re1}
\eeq
where $S_{P}({\bf k})\, (\,S_{P}(x)\,)$ denotes the amplitude to produce
a single particle with the momentum ${\bf k}$ ( at the space-time
coordinate $x$ ) from the initial particle of momentum ${\bf P}$, and
the 3-dimensional product ${\bf kx}$ reads~:
%%%%% re2
\beq
{\bf k \, x}= k_{L} \, x_{L} + {\bf k_{T} \, x_{T}};
\lab{re2}
\eeq
where indices L and T denote longitudinal and transverse components of
momenta and coordinates  with respect to the ${\bf P}$ axis.

Since DLA amplitude $S_{P}=S_{P}(\omega,\theta_{Pk})$ (\ref{4})
depends explicitly on $\omega=\mid{\bf k}\mid\equiv k$ (\ref{9}) and
$\theta_{Pk}$, one should express it as a function of $k_{L}$
and ${\bf k}_{T}$. To proceed, let us recall that the simple form of
$S_{P}=S_{P}(k,\theta_{Pk})$ has been obtained from the original QCD expression
using the approximation of small emission angles $\theta_{Pk}\ll1$ \cite{doksh}.
In this approximation one retains only the leading term in $\theta_{Pk}$.
Hence,
restricting ourselves to the leading terms in $\theta_{Pk}$, we interprete
product
$k\theta_{Pk}$ as the transverse momentum $k_{T}$ and $k$ as the longitudinal
momentum $k_{L}$.
The correctness of the substitution $k\approx k_{L}$ for $e^{ikt}$ in
(\ref{re1})
is not so evident,
since it should be done in the argument of Fourier transform. However, if we
restrict ourselves to finite, small time intervals,
i.\ e.\ $t\ll(P\theta^{2})^{-1}$,
it also works with a high level of accuracy. To see this, let us analyse the
dependence of $k$ on $k_{L}$ and $k_{T}$.
Following (\ref{9}), it can be expressed as~:
%%%% re3
\beq
(\,\omega\equiv\,)k=\sqrt{k_{L}^{2}+k_{T}^{2}}.
\lab{re3}
\eeq
For intrajet cascades there is a relation $k_{T}\ll \, k_{L}$. Restricting
ourselves to the leading term $k\approx k_{L}$, we ignore the contributions
of the order of
$\frac{k_{T}^{2}}{k_{L}}$. In DLA the ratio $\frac{k_{T}^{2}}{k_{L}^{2}}$
has a limiting maximum value
$\theta^{2}$ denoting the emission angle of the previous parton splitting
off the $P$-line (compare (\ref{6})). Maximal energy carried by the gluon
equals $P$. Since that the approximation $k\approx k_{L}$
implies, that one neglects terms of the order of $P\theta^{2}$.

Using the leading term approximation,
Eq.\ (\ref{re1}) can be rewritten as the product of 2 separate Fourier
transforms: 1-dimensional FT of $(k_{L}, t -x_{L})$, and 2-dimensional FT of
$({\bf k_{T}},-{\bf x_{T}})$ in the form~:
%%
%%% re5
\beq
S_{P}(x_{L},{\bf x_{T}},t)= \int dk_{L}\, e^{ik_{L}(t-x_{L})}
\int d^{2}k_{T}\,e^{ -i{\bf k_{T} x_{T}} }
S_{P}(k\approx k_{L},{\bf k_{T}}).
\lab{re5}
\eeq
Eq.\ (\ref{re5}) gives us a good tool to investigate the space-time structure
of DLA.
We may now concentrate on particle distribution in transverse coordinates,
which is physically the most interesting case.
For the sake of simplicity, let us consider for $t=0$ the single particle
amplitude
$S_{P}(k,{\bf x_{T}})$ in mixed transverse space and longitudinal
momentum coordinates \cite{mueller} defined as~:
%%%%% re55
\beq
S_{P}(k_{L}\approx k,{\bf x_{T}},0)=
\int d^{2}k_{T}\, e^{- i{\bf k_{T} x_{T}} }
S_{P}(k_{L}\approx k,{\bf k_{T}}).
\lab{re55}
\eeq
For the single particle exclusive distribution $\rho_{P}^{ex}(k,{\bf x_{T}},0)$,
defined as the amplitude square~:
%%%%% re6
\beq
\rho_{P}^{ex}(k,{\bf x_{T}},0)\,=\,\mid S_{P}(k,{\bf x_{T}},0) \mid^{2};
\lab{re6}
\eeq
one obtains simple expression in the form~:
%%%%% re7
\bea
\rho_{P}^{ex}(k,{\bf x_{T}},0)
 & = & \frac{1}{(2\pi)^{2}}\int d^{2}k_{T}d^{2}k'_{T}
e^{-i({\bf k_{T}}-{\bf k'_{T}}){\bf x}_{T} }
S_{P}(k,{\bf k_{T}})S_{P}^{*}(k,{\bf k'_{T}})\nonumber\\
 & = & \frac{1}{(2\pi)^{2}}\int d^{2}k_{T}d^{2}k'_{T}
e^{-i({\bf k_{T}}-{\bf k'_{T}}){\bf x}_{T} }
d_{P}^{ex}(k,{\bf k'_{T}}; k,{\bf k_{T}}).
\lab{re7}
\eea
%
%which contains as expected only Fourier transform in transverse coordinates.
Relation (\ref{re7}) holds also for inclusive single particle
density (for proof see Appendix G)~:
%%%%% re8
\bea
\rho_{P}^{in}(k,{\bf x_{T}},0)
 & = & \frac{1}{(2\pi)^{2}}\int d^{2}k_{T}d^{2}k'_{T}\,
e^{-i({\bf k_{T}}-{\bf k'_{T}}){\bf x}_{T} }
d_{P}^{in}(k,{\bf k'_{T}}; k,{\bf k_{T}});
\lab{re8}
\eea
and can be easily generalized for any multiparticle
distributions. In this paper we study the inclusive single
particle density in mixed coordinates $\rho_{P}^{in}(k,{\bf x_{T}},0)$
for the QCD gluonic cascade in DLA approximation.
%
%
%                              4
\section{Inclusive single particle density in configuration space}
%                              4.1
%
\subsection{Physics}
Momentum and configuration space description of a particle source are related
to each
other by the Fourier transform. Therefore one expects to extract from the
observables in momentum space some qualitative information about their
behaviour
in configuration space. This observation applies, of course, to single
particle density.
In the DLA approach the inclusive single particle density
$\rho_{P}^{in}({\bf k})$ \cite{wos} reads~:
%%%%p1
\bea
\rho_{P}^{in}({\bf k})&=&
\frac{2b}{2\pi}\frac{1}{kk_{T}^{ 2}} \sum_{n=0}^{\infty}
\frac{(2b)^{n}}{(n!)^{2}}
ln^{n}(\frac{P}{k})
ln^{n}(\frac{k_{T}}{\qu})\,
\Theta_{P}(k)\nonumber\\
& & \nonu\\
&=& \frac{2b}{2\pi}\frac{1}{kk_{T}^{ 2}}
I_{0}\left(\sqrt{8b\,ln(\frac{P}{k})ln(\frac{k_{T}}{\qu})}\right)
\,\Theta_{P}(k);
\lab{p1}
\eea
where $b\equiv g_{S}^{2} C_{F}$.
For constant $k$ the $\rho_{P}^{in}({\bf k})$ is concentrated around {\bf P}
direction (see Fig.\ 6)( skipping for now the $\Theta_{P}(k)$ limitations ).
We expect, that the single particle distribution $\rho_{P}^{in}(t,{\bf x})$
in configuration space~:
%%%%%p2
\beq
\rho^{in}_{P}(k,{\bf x_{T}},0)=
\frac{1}{(2\pi)^{2}}\int d^{2}k_{T}d^{2}k'_{T}
\,e^{-i({\bf k_{T}}-{\bf k'_{T}}){\bf x}_{T} }
\,d^{in}_{P}(k,{\bf k'_{T}};k,{\bf k_{T}});
\lab{p2}
\eeq
will be concentrated around the P direction, as well. However,
it has to obey the uncertainty principle. Since there are
cut-offs of form (\ref{6}) in density matrix, its Fourier transform
will contain some oscillations, due to the restricted integration region.
Moreover, relation (\ref{dm6}) implies, that both densities
$\rho_{P}^{in}({\bf k})$
and $\rho^{in}_{P}(k,{\bf x_{T}},0)$ have the same normalization,
if integrated over $d^{3}k$ and $dk\,d^{2}x_{T}$ respectively.

Remembering the above remarks, we propose now a technique for deriving
the explicit form of inclusive single particle density in configuration space.
%
%                               4.2
\subsection{Single particle density}
We may calculate the single particle density $\rho_{P}^{in}(k,{\bf x_{T}},0)$
as a Fourier transform of DLA density matrix (\ref{d3}).
Let us make the transverse Fourier transform of both sides of Eq.\ (\ref{d3}).
We obtain~:
%%%%%%%ie1
\bea
& & \rho_{P}^{in}(k,{\bf x_{T(P)}},0)=\frac{1}{(2\pi)^{2}}
\int d^{2}k_{T(P)}d^{2}k_{T(P)}'
e^{-i({\bf k}_{T(P)}-{\bf k}_{T(P)}'){\bf x}_{T(P)}}\,
d^{in}_{P}(k,{\bf k'_{T}};k,{\bf k_{T}})\nonu\\
& & \nonu\\
& & \stackrel{\rm DLA}{=}\frac{1}{(2\pi)^{2}}
\int d^{2}k_{T(P)}d^{2}k_{T(P)}'
e^{-i({\bf k}_{T(P)}-{\bf k}_{T(P)}'){\bf x}_{T(P)}}\times\nonumber\\
& & \nonu\\
& & \times\sum_{n=0}^{\infty} (2b)^{n}
\int \frac{ds_{1}}{s_{1}}
\int \frac{d\Omega_{s_{1}}}{2\pi\theta_{P1}^{2}} \Theta_{P}(s_{1}).\,.
\int \frac{ds_{n}}{s_{n}}
\int
\frac{d\Omega_{s_{n}}}{2\pi\theta_{n-1,n}^{2}}\Theta_{s_{n-1}}(s_{n})\nonu\\
& & \nonu\\
& & \times\,g_{s_{n}}^{(1)}(k,{\bf k_{T(S)}'};k,{\bf k_{T(S)}});
\lab{ie1}
\eea
where indices {\small T(P),T(S)} correspond to the reference frame with
the $z-$axis placed along the $P\,(S)$ direction.
The term of the lowest order ($n=0$) takes then the form~:
%%%%%% d4
\bea
&&\rho_{P}^{in\,(1)}(k,{\bf x_{T(P)}},0)=\nonumber\\
& & \nonu\\
&& =\frac{1}{(2\pi)^{2}}
\int d^{2}k_{T(P)}d^{2}k_{T(P)}'e^{-i({\bf k_{T(P)}-k_{T(P)}'}){\bf x_{T(P)}}}
g_{P}^{(1)}(k,{\bf k_{T(S)}'};k,{\bf k_{T(S)}})\nonumber\\
& & \nonu\\
&& =\frac{2b}{2\pi} \, k \,\Theta(\frac{\qu}{\theta}<k<P)
(\int_{\frac{\qu}{k}}^{\theta} d\theta_{Pk} J_{0}(x_{T}k\theta_{Pk}))^{2}.
\lab{d4}
\eea
The numerical plot of (\ref{d4}) is presented in Fig.\ 7. There is a limiting
maximum
value for $x_{T}=0$, as expected, and for large $x_{T}$ the function
has a power decrease with the best fit exponent $x_{T}^{-3.07}$.
The result confirms our intuitive analysis from section
4.\ 1. The term $\rho_{P}^{in\,(1)}(k,{\bf x_{T(P)}},0)$ is concentrated around
the {\bf P} direction, however stronger than
$\rho_{P}^{in\,(1)}(k,{\bf k_{T}})$ from Eq.\ (\ref{p1}).

Derivation of terms of an arbitrary order in expansion (\ref{d3}) is more
complicated. To proceed, let us analyse the last part of (\ref{d3}).
It reads~:
%%%%% d5
\bea
&  & \int \frac{ds_{n}}{s_{n}}\int \frac{d\Omega_{n}}{2\pi\theta_{n-1,n}^{2}}
\, \Theta_{n-1}(s_{n})
\, g_{n}^{(1)}(k,{\bf k_{T(S)}'};k,{\bf k_{T(S)}})=\nonumber\\
& & \nonu\\
& & \nonu\\
&  & =\frac{2b}{2\pi k^{3}} \int \frac{ds_{n}}{s_{n}}
\int \frac{d\Omega_{n}}{2\pi\theta_{n-1,n}^{2}}\, \Theta_{n-1}(s_{n})
\frac{1}{\theta_{nk'}\theta_{nk}}\, \Theta_{n}(k')\Theta_{n}(k);
\lab{d5}
\eea
where the index $(i)$ refers to the vector ${\bf s}_{i}$. Taking into account
phase space
restriction and dominating contribution of
$\frac{1}{\theta_{nk'}\theta_{nk } }$
(see Appendix E) expression (\ref{d5}) can be approximated as~:
%%%%%% d6
\bea
& & \frac{2b}{2\pi k^{3}}\int \frac{ds_{n}}{s_{n}}
\int\frac{d\Omega_{n}}{2\pi\theta_{n-1,n}^{2}}\, \Theta_{n-1}(s_{n})
\frac{1}{\theta_{nk'}\theta_{nk}}
\, \Theta_{n}(k')\Theta_{n}(k)=\nonumber\\
& & \nonu\\
& & =\frac{2b}{2\pi k^{3}}\ln({\frac{s_{n-1}}{k}})\frac{1}{\theta_{n-1,k'}
\theta_{n-1,k}}
\Theta_{n-1}(k')\Theta_{n-1}(k)\times\nonumber\\
& & \nonu\\
& & \times\int_{0}^{\infty}dx_{n}x_{n}J_{0}(x_{n}\theta_{kk'})
\int_{\frac{\qu}{k}}^{\theta_{n-1,k}} d\theta_{nk} J_{0}(x_{n}\theta_{nk})
\int_{\frac{\qu}{k}}^{\theta_{n-1,k'}} d\theta_{nk'} J_{0}(x_{n}\theta_{nk'}).
\lab{d6}
\eea
Scheme (\ref{d6}) can be repeated iteratively in (\ref{d3}). After some
transformations
one obtains finally (for details see Appendix E, H )~:
%%%%% d7
\bea
& & \rho_{P}^{in}(k,{\bf x_{T(P)}},0)=\lab{d7}\\
& & \nonu\\
& & =\frac{2b}{2\pi}\, k\, \Theta(\frac{\qu}{\theta}<k<P)\,
\int_{\frac{\qu}{k}}^{\theta} d\theta_{Pk}
\int_{\frac{\qu}{k}}^{\theta} d\theta_{Pk'}
\, \frac{1}{2\pi}\int_{0}^{2\pi}d\varphi_{k'k}\,
J_{0}(k\theta_{k'k}x_{T})\nonumber\\
& & \nonu\\
& & \times I_{0}(\sqrt{8b\, \ln(\frac{P}{k})
\left\{\int_{0}^{\infty}dx x J_{0}(x\theta_{kk'})\right.
\left.\int_{\frac{\qu}{k}}^{\theta_{Pk}} da J_{0}(xa)\right.
\left.\int_{\frac{\qu}{k}}^{\theta_{Pk'}} da' J_{0}(xa')\right\}})=
\nonu\\
& & \nonu\\
& & \nonu\\
& & =\frac{2b}{2\pi}\, k\, \Theta(\frac{\qu}{\theta}<k<P)\,
\int_{\frac{\qu}{k}}^{\theta} d\theta_{Pk}
\int_{\frac{\qu}{k}}^{\theta} d\theta_{Pk'}
\, \frac{1}{2\pi}\int_{0}^{\infty}
\frac{d\theta_{k'k}\,\theta_{k'k}}
{\Delta(\theta_{Pk},\theta_{Pk'},\theta_{k'k})}
J_{0}(k\theta_{k'k}x_{T})\nonumber\\
& & \nonu\\
& & \times I_{0}(\sqrt{8b\, \ln(\frac{P}{k})
\left\{\right.
\left.\int_{\frac{\qu}{k}}^{\theta_{Pk}} da \right.
\left.\int_{\frac{\qu}{k}}^{\theta_{Pk'}} da'\right.
\left.\frac{1}{2\pi\Delta(a,a',\theta_{k'k})}\right\}
});\nonu
\eea
where $\theta_{k'k}$ denotes the angle between momenta ${\bf k'}$,
${\bf k}$, which in the reference frame with the z-axis placed along
P direction takes the simple form~:
%%%%%% d8
\beq
\theta_{k'k}=\sqrt{\theta_{Pk'}^{2}+\theta_{Pk}^{2}
-2\,\theta_{Pk'}\theta_{Pk} \cos(\varphi_{k'k})};
\lab{d8}
\eeq
$\varphi_{k'k}$ denotes the azimuthal angle between ${\bf k_{T}}$
and ${\bf k'_{T}}$, and $\Delta(a,b,c)$ denotes the area of the triangle
with sides $a$, $b$, $c$.
The numerical plot of (\ref{d7}) is presented in Fig.\ 8~. There is a limiting
maximum
value for $x_{T}=0$, as expected, and for large $x_{T}$ the function
has a power decrease with the best fit exponent $x_{T}^{-2.43}$.
The result confirms our intuitive analysis from section
4.\ 1. The term $\rho_{P}^{in\,(1)}(k,{\bf x_{T(P)}},0)$ is concentrated around
the {\bf P} direction, stronger than
$\rho_{P}^{in\,(1)}(k,{\bf k_{T}})$ from Eq.\ (\ref{p1}). The function
oscillates around its power-law profile. As already mentioned in section
4.\ 1.\ , this effect is due to the sharp cut-offs of form (\ref{6}) which
restrict the integration space.

%                           4.3
\subsection{Properties}
Let us analyse the properties of (\ref{d7}) in detail. Comparing (\ref{d7})
with the single particle density in momentum space (\ref{p1}), we get the same
dependence on the energy k, as expected. The complicated emission term
(see Fig.\ 9)~:
%%%%% factor
\bea
& & 2b\,\ln\left(\frac{P}{k}\right)\,
\left\{\int_{0}^{\infty}dx x J_{0}(x\theta_{kk'})\right.
\left.\int_{\frac{\qu}{k}}^{\theta_{Pk}} da J_{0}(xa)\right.
\left.\int_{\frac{\qu}{k}}^{\theta_{Pk'}} da' J_{0}(xa')\right\}=\nonumber\\
& & =\,\int\, d^{3}s \, \frac{k^{3}}{s^{3}}\,A_{ss}(k,{\bf k'}_{T};
k,{\bf k}_{T})
\lab{factor}
\eea
corresponds to the logarithm $\ln\frac{k_{T}}{\qu}$ of (\ref{p1}).
In fact, after integration of $\rho_{P}^{in}(k,{\bf x_{T(P)}},0)$
over $d^{2}x_{T}$
one obtains the same result  as for (\ref{p1}) integrated over $d^{2}k_{T}$
(see Appendix I).
The density $\rho_{P}^{in}(k,{\bf x_{T(P)}},0)$ is positively defined, as well.
Using the Bessel function identities from Appendix H,
Eq.\ (\ref{d7}) can be rewritten in the form~:
%%%%%% d9
\bea
& & \rho_{P}^{in}(k,{\bf x_{T(P)}},0)=\lab{pm1}\\
& & \nonu\\
& & =\frac{2b}{2\pi}\, k \Theta(\frac{\qu}{\theta}<k<P)\,
\sum_{n=0}^{\infty} \frac{(2b)^{n}}{(n!)^{2}}\, \ln^{n}(\frac{P}{k})
(\int_{0}^{\infty}dx_{1} x_{1} .\, .\int_{0}^{\infty}dx_{n} x_{n})
\sum_{m_{1}.\, .m_{n}=-\infty}^{\infty}\nonu\\
& & \nonu\\
& & \times \left\{\int_{\frac{\qu}{k}}^{\theta} d\theta_{Pk}\right.
J_{m_{1}}(x_{1}\theta_{Pk}).\, .J_{m_{n}}(x_{n}\theta_{Pk})\,
J_{m_{1}+\ldots+m_{n}}(x_{T}k\theta_{Pk})
\prod_{i=1}^{n}
\left.\int_{\frac{\qu}{k}}^{\theta_{Pk}} da_{i} J_{0}(x_{i}a_{i})\right\}^{2};
\nonu
\eea
which implies positive definitness.
For $x_{T}=0$ $\rho_{P}^{in}(k,{\bf x_{T(P)}},0)$
reaches its maximal value, as expected. Diagramatically,
formula (\ref{d7}) represents the chain of independent emission factors
(\ref{factor})
transformed onto the transverse $x_{T}$ plane by the Bessel factor
of primary emission from the parent P, namely
$J_{0}(\mid {\bf k}_{T} - {\bf k'}_{T}\mid x_{T})$ (see Fig.\ 10).
For $k=P$ expression (\ref{d7}) reduces to (\ref{d4}) with the dominance of the
primary emission, as expected.

%                            5
\section{Summary}
We considered the QCD parton cascade created in $e^{+}e^{-}$ collision in the
double logarithmic approximation. Using the density matrix (DM) formalism
\cite{ziaja}, and restricting ourselves to the terms leading in the double
logarithmic (DL)
perturbative expansion for the quasi-diagonal limit $k'\cong k$,
we derived the explicit form of the inclusive single
particle density matrix $d_{P}^{in}(k',k)$ and single particle density
$\rho_{P}^{in}(k,{\bf x_{T(P)}},0)$ (see Fig.\ 8) in mixed coordinates
$(k_{L}\approx k,{\bf x}_{T})$. The gluon density
$\rho_{P}^{in}(k,{\bf x_{T(P)}},0)$ fulfills important physical requirements
such as positive definitness and proper normalization. It is concentrated
around the ${\bf P}$ direction, and shows the power law profile for large
${\bf x}_{T}$. Due to the cut-offs of the type (\ref{6}) which restrict
kinematic regions, the density oscillates around its power law profile.

The above results give a positive outlook for the future. The simplified
technique
for calculating multiparticle observables in DLA for the constant $\alpha_{S}$
is ready. It may allow one to investigate the structure of the QCD cascade in a
very comprehensive way: the exact form of a Wigner function already would give
us clear experimental predictions.

In order to improve the approach one needs to include the momentum dependence
of $\alpha_{S}$. The exact analysis of applicability of the quasi-diagonal
approximation \newline $k'\cong k$ would be required, as well.
There is also a problem as to what extent the kinematical constraints
characteristic for a given QCD approach ( in DLA they are of the form
(\ref{6}) ) influence the results obtained in the approach. In other
words, there is a problem how to separate the effects of the kinematical
restriction from the dynamical content of QCD space.
The explanation would require performing an analysis similar to the presented
above for LLA and MLLA schemes, as well. The MLLA approximation would be of
special interest since it would incorporate the energy conservation into
the cascade.
\section{Appendix A}
%%%% strona 40',47'
The functionals  $Z_{P}[u]$ and $Z_{P'}[s]$ defined in (\ref{16}) generate
the sum of amplitudes (\ref{4}) and their complex conjugates over all
possible tree diagrams respectively. It can be seen when one rewrites
the master equation for $Z_{P}$ in the form of the diagram series
(see Fig.\ 2). From that construction follows exclusive density matrix
(\ref{17ex})~:
%%%%% 32
\beq
\roexmp = \frac{\delta^{m}}{\delta s_{1'}\ldots\delta s_{m'}}
Z_{P}^{*}[s]\mid_{\{s=0\}}
\frac{\delta^{m}}{\delta u_{1}\ldots\delta u_{m}} Z_{P}[u]\mid_{\{u=0\}}.
\lab{32}
\eeq
Then the inclusive density matrix calculated from (\ref{dm3}) looks like this~:
%%%%% 33
\bea
\roinmp = \sum_{n=0}^{\infty} \frac{1}{n!} \,
\left( \int d^{3}k \frac{\delta}{\delta s} \frac{\delta}{\delta u}\right)^{n}
\frac{\delta^{m}}{\delta s_{1'}\ldots\delta s_{m'}} Z_{P}^{*}[s]\mid_{\{s=0\}}
\nonu\\
\times\, \, \frac{\delta^{m}}{\delta u_{1}\ldots\delta u_{m}}
Z_{P}[u]\mid_{\{u=0\}}\,\,;
\lab{33}
\eea
and from the identity for the product of any two functionals F, F'~:
%%%%% 34
\beq
F[u]F'[w]\mid_{\{u=\frac{\del}{\del w}, w=0\}}=
\sum_{n=0}^{\infty} \frac{1}{n!} \,
\left( \int d^{3}k \frac{\delta}{\delta w} \frac{\delta}{\delta u}\right)^{n}
F[u]\mid_{\{u=0\}}F'[w]\mid_{\{w=0\}};
\lab{34}
\eeq
follows (\ref{17}).

\section{Appendix B}

%%%% strona75',77'
In DLA there are four different tree graphs $M_{a}, M_{b}, M_{c}, M_{d}$,
describing the production of two gluons.
They are defined on non-overlapping kinematic regions \newline (see Fig.\ 3 ).
Emitted gluons are either angular (AO) or energy ordered (EO).

Let us consider all the diagrams contributing to the single particle density
matrix
$d_{P}^{ex}(k_{1}',k_{1})$. From the (AO) and (EO) it follows that the
interference
between any two different graphs will appear only if either energies $\om_{1},
\om_{1}'$ or emission angles $\theta_{1P}, \theta_{1'P}$ of produced gluons
are {\sl strongly} ordered (Fig.\ 4).

This statement can be generalized for any m-particle density matrix by
induction.
If we have m particles, and (m-1) ones among them are "close" to each other,
i.\ e.\
$ k_{1}\cong k_{1}',\ldots,k_{m-1}\cong k_{m-1}'$, then the interference
of the different diagrams will take  place only if either energies or
angles of $ k_{m}$ and $k_{m}'$ are strongly ordered.

\section{Appendix C}
%%% strony 8.III

$Z_{P'P}[u(k',k)]$ defined in (\ref{dla2}) generates correct exclusive density
matrices (\ref{dla1ex}). The proof of this statement follows from the
representation
of the master equation for $Z_{P'P}$ as a diagram series (see Fig.\ 5).
The series reproduces all squared contributions of identical tree diagrams,
and excludes interference of the different ones
(function ${\em P}_{1',\ldots,n';1,\ldots,n}$).

The explicit form of inclusive density (\ref{dla1}) follows from formula
(\ref{dm3}). Substituting (\ref{dla1ex}) into (\ref{dm3}) one obtains~:
%%%%% 35
\bea
& & \roinmp=\nonu\\
& & \nonu\\
& &=\,\sum_{n=0}^{\infty} \frac{1}{n!} \,
\left( \int \right.
\left. d^{3}k d^{3}k' \del^{3}(k'-k) \frac{\delta}{\delta u_{k',k}}\right)^{n}
\, \frac{\delta^{m}}{ \delta u_{m',m}\ldots\delta u_{1',1}  }
Z_{P'P}[u]\mid_{\{u=0\}};\nonu\\
& &
\lab{35}
\eea
which represents the functional $Z_{P'P}[u]$ expanded around "null" $\{u=0\}$
in the "point" $u=\dl$.

\section{Appendix D}

%%%strony I.74
We want to average relation (\ref{dla4}) over physical polarizations.
The exact expression to be summed over polarizations looks like this~:
%%%%%% 36
\beq
\frac{( e \cdot P) (e' \cdot P')}{(k \cdot P)(k' \cdot P')};
\lab{36}
\eeq

\nin
where $e, e'$ are polarizations of the same gluon identical in the limit $k'=k$.
The gauge fixing we apply in the approach allows us to neglect contributions to
(\ref{36}) coming from the "nonphysical"
polarizations $e^{0}$ and $e^{3}$.
Furthermore, it requests the time components of the physical polarizations
$e^{1}, e^{2}$ to be equal to $0$. Space components of $e^{1}, e^{2}$
can be then constructed in the form~:
%%%%%% 37
\bea
{\bf e}^{2} & = &  \frac{ {\bf P \times k}}{ \mid {\bf P \times k}\mid }\,
,\nonu\\
{\bf e}^{1} & = &  \frac{ {\bf e^{2} \times k}}{ \mid {\bf e^{2} \times k}
\mid };
\lab{37}
\eea

\nin
and for $e'$ respectively~:
%%%%%% 38
\bea
{\bf e}^{\prime 2} & = &  \frac{ {\bf P' \times k'}}{ \mid {\bf P' \times k'}
\mid }
\,,\nonu\\
{\bf e}^{\prime 1} & = &  \frac{ {\bf e^{\prime 2} \times k'}}{ \mid
{\bf e^{\prime 2}\times k'}\mid }.
\eea
Summing over these 2 polarizations and expanding scalar products in (\ref{36}),
one obtains finally~:
%%%%%% 39
\bea
\sum_{j=1,2}\frac{( e^{j} \cdot P) (e^{\prime j} \cdot P')}{(k \cdot P)
(k' \cdot P')} & = &
\frac{4}{\om \om'} \,\, \frac{1}{\theta_{Pk} \theta_{P'k'}}.
\lab{39}
\eea

The result can be easily confirmed for any two physical polarizations
${\bf \ep}^{1}, {\bf \ep}^{2}$ placed on the plane ${\bf e}^{1} {\bf e}^{2}$~:
%%%%% 40
\bea
{\bf \ep}^{1} & = & \,\,\, cos \varphi \, {\bf e}^{1} + sin \varphi \,
{\bf e}^{2}\,,\nonu\\
{\bf \ep}^{2} & = &  -sin \varphi \, {\bf e}^{1} + cos \varphi \, {\bf e}^{2};
\lab{40}
\eea

\nin and for ${\bf \ep}^{\prime 1}, {\bf \ep}^{\prime 2}$ placed on the plane
${\bf e}^{\prime 1}{\bf e}^{\prime 2}$ with the phases
$\varphi'=\varphi$ respectively.
%
%                               5
\section{Appendix E}

To transform (\ref{d5}) into (\ref{d6}) we will apply the pole approximation
method (\cite{wos} and references therein). The phase space restrictions
$\Theta_{n-1}(n)\Theta_{n}(k)\Theta_{n}(k')$ in (\ref{d5}), in particular
the angular ordering (AO)
$ \theta_{n-1,n}\gg\theta_{n,k}, \, \theta_{n-1,n}\gg\theta_{n,k'}$ make
the the term $\frac{1}{\theta_{nk'}\theta_{nk } }$ be dominating in (\ref{d5}).
Applying the Bessel function identity \cite{mueller}~:
%%%%% a1
\bea
\int \frac{d\Omega_{n}}{2\pi}
& = &\frac{1}{2\pi}\, \int \frac{d\theta_{nk}\,d\theta_{nk'}\,
\theta_{nk}\,\theta_{nk'}}{\Delta(\theta_{nk},\theta_{nk'},\theta_{kk'})}\,=
\nonu\\
& & \nonu\\
& = &\int _{0}^{\infty} dx x J_{0}(x\theta_{k'k})
\int d\theta_{nk}\theta_{nk}\, J_{0}(x\theta_{nk})
\int d\theta_{nk'}\theta_{nk'}\, J_{0}(x\theta_{nk'});
\lab{a1}
\eea
where $\theta_{k'k}$ denotes the relative angle between vectors
${\bf k},{\bf k}'$,
and $\Delta(\theta_{nk},\theta_{nk'},\theta_{kk'})$ equals the area of triangle
with sides $\theta_{k'k},\,\theta_{k'n},\,\theta_{kn}$,
we may rewrite expression (\ref{d5}) in the form~:
%%%%% a2
\bea
& & \frac{2b}{2\pi k^{3}}\int\frac{ds_{n}}{s_{n}}
\int\frac{d\Omega_{n}}{2\pi\theta_{n-1,n}^{2}}\Theta_{n-1}(n)
\frac{1}{\theta_{nk'}\theta_{nk}}
\Theta_{n}(k')\Theta_{n}(k)=\nonumber\\
& & \nonu\\
& & \nonu\\
& & =\frac{2b}{2\pi k^{3}} \int \frac{ds_{n}}{s_{n}}
\int_{0}^{\infty}dx_{n}x_{n}J_{0}(x_{n}\theta_{kk'})
\int d\theta_{nk} J_{0}(x_{n}\theta_{nk})
\int d\theta_{nk'} J_{0}(x_{n}\theta_{nk'})\nonumber\\
& & \nonu\\
& & \hspace*{\fill}
\times\,
\frac{1}{\theta_{n-1,n}^{2}}\Theta_{n-1}(n)\Theta_{n}(k')\Theta_{n}(k).
\lab{a2}
\eea
Because of angular ordering the angle $\theta_{n-1,n}$ practically does not
change while integrating over angles $\theta_{nk},\theta_{nk'}$, and can be
successfully approximated by the angle $\theta_{n-1,k}$ ($\theta_{n-1,k'}$).
Applying all the integration restrictions explicitly, we arrive finally
at the expression~:
%%%%% a3
\bea
\frac{2b}{2\pi k^{3}}\int\frac{ds_{n}}{s_{n}}
\int\frac{d\Omega_{n}}{2\pi\theta_{n-1,n}^{2}}\Theta_{n-1}(n)
\frac{1}{\theta_{nk'}\theta_{nk}}
\Theta_{n}(k')\Theta_{n}(k)\,\,\,\,=\nonumber\\
\nonu
\eea
\bea
&  & =\frac{2b}{2\pi k^{3}}\frac{1}{\theta_{n-1,k'}\theta_{n-1,k}}
\Theta_{n-1}(k')\Theta_{n-1}(k) \int_{k}^{s_{n-1}} \frac{ds_{n}}{s_{n}}
\times\\
& & \nonu\\
& & \times\int_{0}^{\infty}dx_{n}x_{n}J_{0}(x_{n}\theta_{kk'})
\int_{\frac{\qu}{k}}^{\theta_{n-1,k}} d\theta_{nk} J_{0}(x_{n}\theta_{nk})
\int_{\frac{\qu}{k}}^{\theta_{n-1,k'}} d\theta_{nk'} J_{0}(x_{n}\theta_{nk'});
\nonu
\lab{a3}
\eea
which after trivial integration over $s_{n}$ gives identity (\ref{d6}).
%
%                                6
\section{Appendix F}
We shall find such terms in expansion (\ref{ie2}) which dominate
in the double logarithmic perturbative limit of $d_{P}^{in}(k',k)$,
i.\ e.\ when $b\,\rightarrow 0$ ($b\propto\alpha_{S}$) and $P\rightarrow\infty$,
so that $b \ln^{2}\frac{P}{\qu}$ remains constant, and generate double
logarithmic
corrections to the cross section.
First let us introduce the following notation~:
%%%% now0
\bea
d_{P}^{in}(k',k)&=& \sum_{n=0}^{\infty}\, a_{n};
\lab{now0}
\eea
where coefficients $a_{n}$ describe the nth order iteration of
$g_{S}(k',k)$ (\ref{ie2}) in the form~:
%%%% now1
\bea
a_{n}\,=\,(2b)^{n}
\int \frac{ds_{1}}{s_{1}}
\int \frac{d\Omega_{s_{1}}}{2\pi\theta_{Ps1}^{2}} \Theta_{P}(s_{1}).\,.
\int \frac{ds_{n}}{s_{n}}
\int
\frac{d\Omega_{s_{n}}}{2\pi\theta_{n-1,n}^{2}}\Theta_{s_{n-1}}(s_{n})
\,g_{s_{n}}(k',k).
\lab{now1}
\eea
The term $a_{0}$ of the series (\ref{now1}) is equal to $g_{P}(k',k)$.
Expanding it in the powers of coupling constant $b$,
one obtains the term of the first (lowest) order of b in the form~:
%%%% now2
\beq
a_{0}^{(1)}\,\equiv\,g_{P}^{(1)}(k',k)=
\frac{2b}{2\pi}\,\frac{1}{\sqrt{k^{3}k^{\prime 3}}}\,
\frac{1}{\theta_{Pk'}\theta_{Pk}}
\Theta_{P}(k')\, \Theta_{P}(k).
\lab{now2}
\eeq
Denoting the other terms by the symbol
$a_{0}^{(>1)}\,\equiv\,g_{P}^{(>1)}(k',k)$,
the term $a_{0}$ from the series (\ref{now1}) can be rewritten as~:
%%%% now22
\beq
a_{0}=a_{0}^{(1)}\,+\,a_{0}^{(>1)}.
\lab{now22}
\eeq
We have checked that $a_{0}^{(1)}$ dominates in (\ref{now22}). To see this,
one should rewrite $a_{0}$ as~:
%%%%% now11
\bea
a_{0}\,\equiv\,g_{P}(k',k)\,& = &\,a_{0}^{(1)}\,+\,a_{0}^{(>1)}\\
                            & = &\,a_{0}^{(1)}\,f(k',k ; P);
\lab{now11}
\eea
where the exact form of $f(k',k; P)$ follows from (\ref{ie3}).
For $b\,>\,0$ and $P$ finite the $f(k',k ; P)$ vs $\mid {\bf k-k'}\mid$
is a gauss-like function with the maximum equal to $1$ for ${\bf k}={\bf k}'$
and a non-zero minimal value. In the DL perturbative limit~:
$f(k', k; P) \stackrel{\rm PT}{\rightarrow}1$. Consequently, for any $k$, $k'$,
$P$~:
%%%%% now3
\bea
a_{0}\,&\leq & \,a_{0}^{(1)},\lab{now31}\\
& & \nonu\\
\frac{a_{0}}{a_{0}^{(1)}} & \stackrel{\rm PT}{\rightarrow} & 1.
\lab{now32}
\eea
Since $a_{0}^{(1)}$ dominates in $a_{0}$, we expect that the iterations of
$a_{0}^{(1)}$ will generate the leading contributions to the $d_{P}^{in}(k',k)$.
Let us introduce the notation~:
%%%%% now66
\bea
d_{P}^{in,\,(1)}(k',k) & = & \sum_{n=0}^{\infty}\, a_{n}^{(1)},\\
d_{P}^{in,\,(>1)}(k',k)& = & \sum_{n=0}^{\infty}\, a_{n}^{(>1)};
\lab{now66}
\eea
where $a_{n}^{(>1)}$ and $a_{n}^{(1)}$ represent the result
of nth iteration (\ref{now1}) of $a_{0}^{(>1)}$ and $a_{0}^{(1)}$
respectively. From (\ref{now31}) immediately follows the relation~:
%%%% now4
\beq
a_{n}\,=\,a_{n}^{(1)}\,+\,a_{n}^{(>1)}\,\leq \,a_{n}^{(1)}.
\lab{now4}
\eeq
Since $a_{n}\,\geq\,0$, for the density matrix one obtains~:
%%%% now5
\beq
0\,\leq\,
d_{P}^{in}(k',k)\,=\,
d_{P}^{in,\,(1)}(k',k)\,+\, d_{P}^{in,\,(>1)}(k',k)\,
\leq\,d_{P}^{in,\,(1)}(k',k).
\lab{now5}
\eeq
In the limit $\mid {\bf k}-{\bf k'} \mid \,\rightarrow\,0$
the density matrix $d_{P}^{in}(k',k)\,\rightarrow\, d_{P}^{in,\,(1)}(k',k)$,
and therefore~:
%%%% now6
\beq
\frac{d_{P}^{in}(k',k)}{d_{P}^{in,(1)}(k',k)}\,\stackrel{\rm PT}{\rightarrow}
\,1.
\lab{now6}
\eeq
For other ( larger ) values of $\mid {\bf k}-{\bf k}' \mid$ the contribution of
$d_{P}^{in,(1)}(k',k)$
generally need not to dominate. However, taking into account the fact that the
density matrix approach works only for the quasi-diagonal limit,
we may identify the quasi-diagonal region with the region of the dominance of
$d_{P}^{in,(1)}(k',k)$. Furthermore, since $d_{P}^{in,(>1)}(k,k)=0$,
the $d_{P}^{in,(1)}(k',k)$ generates all DL corrections to the cross section.
Hence we finally arrive at the relation (\ref{d3}).
%
%                                7
\section{Appendix G}
The particle density $\rho_{P}^{in}(k,{\bf x}_{T},0)$ can be represented
as~:
\beq
\rho_{P}^{in}(k,{\bf x}_{T},0)=\sum_{n=1}^{\infty} \frac{1}{(n-1)!}
\int dk_{2}d^{2}x_{2T}\ldots dk_{n}d^{2}x_{nT}
\,\rho_{P}^{ex}(k,{\bf x}_{T},k_{2},{\bf x}_{2T},k_{n},{\bf x}_{nT}).
\lab{aa1}
\eeq
Representing the exclusive particle density as the square of multiparticle
amplitudes,
expressing the configuration space amplitudes as Fourier transforms of
amplitudes
in momentum space, and performing integration over $d^{2}x_{2T}\ldots
d^{2}x_{nT}$ finally one arrives at Eq.\ (\ref{re8}).

%                                8
\section{Appendix H}
For our purposes let us quote the following identities for Bessel functions
(\cite{jackson} and references therein)~:
%%%% b1
\bea
J_{0}(x\theta_{k'k}) & = & \sum_{m=-\infty}^{\infty} e^{im(\varphi-\varphi')}
J_{m}(x\theta_{kP})J_{m}(x\theta_{k'P}),\lab{b1}\\
%
%
%%%%% b2
e^{i\,k\,x\, \cos{\varphi}} & = &
\sum_{m=-\infty}^{\infty} i^{m}e^{im\varphi}J_{m}(k\, x),\lab{b2}\\
%
%%%%% b3
\int_{0}^{\infty} dx\, x J_{0}(x\,a)J_{0}(x\,a') & = & \frac{\delta(a-a')}{a};
\lab{b3}
\eea
where $\theta_{k'k}$ denotes the relative angle between momenta ${\bf k'}$
${\bf k}$, which in the reference frame with the z-axis placed along
P direction takes form (\ref{d8}), and $\varphi_{k'k}=\varphi-\varphi'$,
where the angles $\varphi$, $\varphi'$ denote the azimuthal angles of vectors
${\bf k_{T}}$ and ${\bf k'_{T}}$ on the transversal plane respectively.

Now let us repeat the scheme Eq.(\ref{d5})$\rightarrow$ Eq.(\ref{d6})
iteratively in (\ref{d3}). Then, after introducing the explicit form of
Fourier transform we arrive at the expression~:
%%%%% b4
\bea
& & \rho_{P}^{in}(k,{\bf x_{T(P)}},0)=\lab{b4}\\
& & \nonu\\
& & =\frac{1}{(2\pi)^{2}}\int d(k\theta_{kP})\, k\theta_{kP}
                    \int d(k\theta_{kP})\, k\theta_{kP}
                    \int_{0}^{2\pi} d\varphi
                    \int_{0}^{2\pi} d\varphi'
e^{-ik\theta_{kP}x_{T}\cos{\varphi}+ik\theta_{k'P}x_{T}\cos{\varphi'}}
\nonumber\\
& & \nonu\\
& & \times\sum_{n=0}^{\infty} \frac{(2b)^{n}}{n!}\,\ln^{n}
\left(\frac{P}{k}\right)
\frac{2b}{2\pi k^{3}} \frac{1}{\theta_{Pk}\theta_{Pk'}}
                      \Theta_{P}(k)\Theta_{P}(k')\nonu\\
& & \nonu\\
& & \times
\left(\int_{0}^{\infty}dx_{1} x_{1} J_{0}(x_{1}\theta_{k'k}).\, .\right.
\left. \int_{0}^{\infty}dx_{n} x_{n}  J_{0}(x_{n}\theta_{k'k})\right)
\nonumber
\eea
\bea
\times\left\{ \int_{\frac{\qu}{k}}^{\theta_{Pk}} da_{1} J_{0}(x_{1}a_{1})\right.
\int_{\frac{\qu}{k}}^{\theta_{Pk'}}da_{1}'J_{0}(x_{1}a_{1'}) .\,.
\int_{\frac{\qu}{k}}^{\theta_{n-1,k}} da_{n} J_{0}(x_{n}a_{n})
\left. \int_{\frac{\qu}{k}}^{\theta_{n-1,k'}}da_{n}'J_{0}(x_{n}a_{n}')\right\}.
\nonu
\eea
It looks still quite complicated, however one can notice that the term in
brackets~:
\beq
\left\{ \int_{\frac{\qu}{k}}^{\theta_{Pk}} da_{1} J_{0}(x_{1}a_{1})\right.
\int_{\frac{\qu}{k}}^{\theta_{Pk'}}da_{1}'J_{0}(x_{1}a_{1'}) .\,.
\int_{\frac{\qu}{k}}^{\theta_{n-1,k}} da_{n} J_{0}(x_{n}a_{n})
\left. \int_{\frac{\qu}{k}}^{\theta_{n-1,k'}}da_{n}'J_{0}(x_{n}a_{n}')\right\};
\eeq
which contains integration of 2n Bessel functions,
simplifies significantly if one adds to (\ref{b4}) some extra off-diagonal
terms, which are equal to $0$ in the quasi-diagonal limit. Namely~:
%%%%% b5
\beq
\left\{ \, \, \right\} \stackrel{\rm DLA}{=}
\frac{1}{n!}
\int_{\frac{\qu}{k}}^{\theta_{Pk}} da_{1} J_{0}(x_{1}a_{1})
\int_{\frac{\qu}{k}}^{\theta_{Pk'}}da_{1}'J_{0}(x_{1}a_{1'}) .\, .
\int_{\frac{\qu}{k}}^{\theta_{Pk}} da_{n} J_{0}(x_{n}a_{n})
\int_{\frac{\qu}{k}}^{\theta_{Pk'}}da_{n}'J_{0}(x_{n}a_{n}').
\lab{b5}
\eeq
Then expression (\ref{b4}) takes the simpler form~:
%%%%% b6
\bea
& & \rho_{P}^{in}(k,{\bf x_{T(P)}},0)= \frac{2b}{2\pi}\,k \,
\Theta(\frac{Q_{0}}{\theta}<k<P)\frac{1}{(2\pi)^{2}}\times\nonumber\\
& & \nonu\\
& & \times\int_{\frac{\qu}{k}}^{\theta} d\theta_{kP}\,
                    \int_{\frac{\qu}{k}}^{\theta} d\theta_{k'P}\,
                    \int_{0}^{2\pi} d\varphi
                    \int_{0}^{2\pi} d\varphi'
e^{-ik\theta_{kP}x_{T}\cos{\varphi}+ik\theta_{k'P}x_{T}\cos{\varphi'}}\\
& & \nonu\\
& & \times\sum_{n=0}^{\infty} \frac{(2b)^{n}}{(n!)^{2}}\,\ln^{n}
\left(\frac{P}{k}\right)
\left\{\int_{0}^{\infty}dx x J_{0}(x \theta_{k'k}) \right.
\left .\int_{\frac{\qu}{k}}^{\theta_{Pk}} da J_{0}(x a) \right.
\left.\int_{\frac{\qu}{k}}^{\theta_{Pk'}} da' J_{0}(x a') \right\}^{n}.\nonu
\lab{b6}
\eea
Let us perform the integration over the angles $\varphi, \varphi'$.
Replacing all
the terms of the type $J_{0}(x\theta_{k'k})$ by their expansions (\ref{b1}),
expanding then Fourier transforms in terms of (\ref{b2}), and performing
the integration over $\varphi, \varphi'$ explicitly, we arrive at (\ref{pm1}).

However, the part of expression (\ref{pm1}) containing Bessel functions
can be rewritten as~:
%%%%%%% b7
\bea
& & \left( \prod_{i=1}^{n} \sum_{m_{i}=-\infty}^{\infty}
J_{m_{i}}(x_{i}\theta_{Pk}) J_{m_{i}}(x_{i}\theta_{Pk'})\right)
J_{m_{1}+.\, .+ m_{n}}(x_{T}k\theta_{Pk})
J_{m_{1}+.\, .+ m_{n}}(x_{T}k\theta_{Pk'})=\nonumber\\
& & \nonu\\
& & =\frac{1}{2\pi}\int_{0}^{2\pi}d\varphi_{k'k} \left( \prod_{i=1}^{n}
J_{0}(x_{i}\theta_{k'k})\right)
J_{0}(x_{T}k\theta_{k'k});
\lab{b7}
\eea
where $\varphi_{k'k}$ denotes the relative angle between the vectors
${\bf k}_{T},{\bf k}_{T}'$ and \newline
$\theta_{k'k}=\sqrt{\theta_{Pk'}^{2}+\theta_{Pk}^{2}
-2\,\theta_{Pk'}\theta_{Pk} \cos(\varphi_{k'k})}$, as usual. Substituting
(\ref{b7}) into (\ref{pm1}) finally we obtain (\ref{d7}).
%                                 9
\section{Appendix I}
Let us investigate normalization of single particle density (\ref{d7}) in real
space, i.\ e.\ the integral $\int d^{2}x_{T} \rho_{P}^{in}(k,{\bf x}_{T},0)$.
It reads~:
%%%%%%% ad1
\bea
& &\int d^{2}x_{T} \rho_{P}^{in}(k,{\bf x}_{T},0)=\nonumber\\
& & \nonu\\
& &=2bk \Theta(\frac{\qu}{\theta}<k<P)\,
\left(\int_{\frac{\qu}{k}}^{\theta} d\theta_{Pk}
\int_{\frac{\qu}{k}}^{\theta} d\theta_{Pk'}
\frac{1}{2\pi}\int_{0}^{2\pi}d\varphi_{k'k}\right)
\int_{0}^{\infty}dx_{T}x_{T}J_{0}(k\theta_{k'k}x_{T})\nonumber\\
& & \nonu\\
& & \,\,\,\,\times I_{0}(\sqrt{8b\, \ln(\frac{P}{k})
\left\{\int_{0}^{\infty}dx x J_{0}(x\theta_{kk'})\right.
\left.\int_{\frac{\qu}{k}}^{\theta_{Pk}} da J_{0}(xa)\right.
\left.\int_{\frac{\qu}{k}}^{\theta_{Pk'}} da' J_{0}(xa')\right\}}).
\lab{ad1}
\eea
The term $J_{0}(k\theta_{k'k}x_{T})$ is the only part of expression
(\ref{ad1}) depending on $x_{T}$. Integral $\int_{0}^{\infty}dx_{T}
x_{T}J_{0}(k\theta_{k'k}x_{T})$
can be performed, using identities (\ref{b1}),(\ref{b3}). It equals~:
%%%%% ad2
\bea
\int_{0}^{\infty}dx_{T}x_{T}J_{0}(k\theta_{k'k}x_{T}) & = &
2\pi\, \delta(\varphi_{k'k}) \frac{\delta(\theta_{kP}-\theta_{k'P})}{k^{2}
\theta_{kP}}.
\lab{ad2}
\eea
Substituting (\ref{ad2}) into (\ref{ad1}), and integrating over
$\theta_{k'P}$ and $\varphi_{k'k}$, one arrives at~:
%%%%% ad3
\bea
& & \int d^{2}x_{T} \rho_{P}^{in}(k,{\bf x}_{T},0)=
\frac{2b}{k}\,\Theta(\frac{\qu}{\theta}<k<P)
\int_{\frac{\qu}{k}}^{\theta} \frac{d\theta_{Pk}}{\theta_{Pk}}\times\nonumber\\
& & \nonu\\
& & \times I_{0}(\sqrt{8b\, \ln(\frac{P}{k})
\left\{\int_{0}^{\infty}dx x \right.
\left.\int_{\frac{\qu}{k}}^{\theta_{Pk}} da J_{0}(xa)\right.
\left.\int_{\frac{\qu}{k}}^{\theta_{Pk}} da' J_{0}(xa')\right\}}).
\lab{ad3}
\eea
Performing the integration over $x$ in the argument of function $I_{0}$
with the help of identity (\ref{b3}), finally one obtains~:
%%%%% ad4
\beq
\int d^{2}x_{T}\, \rho_{P}^{in}(k,{\bf x}_{T},0)
=\frac{2b}{k}\Theta(\frac{\qu}{\theta}<k<P)
\int_{\frac{\qu}{k}}^{\theta} \frac{d\theta_{Pk}}{\theta_{Pk}}
I_{0}(\sqrt{8b\, \ln(\frac{P}{k}) \ln(\frac{k\theta_{Pk}}{\qu})});
\lab{ad4}
\eeq
which equals the normalization of particle density in momentum space
defined in (\ref{p1})~:
%%%%% ad5
\beq
\int d^{2}x_{T}\,\, \rho_{P}^{in}(k,{\bf x}_{T},0)=
\int d^{2}k_{T}\,\, \rho_{P}^{in}(k,{\bf k}_{T}).
\lab{ad5}
\eeq
%
%\newpage
\vspace{0.5cm}
\bec
{\bf Acknowledgements}
\eec
\vspace{0.5cm}

I would like to express my deep gratitude to Professor Andrzej Bialas
for the  encouragement to study the subject of this work, for many helpful
discussions and suggestions, and a continuous interest and support throughout
this work.
I would like to thank Professor Jacek Wosiek for many supportive and
enlightening discussions and comments, for careful reading the manuscripts,
and for his continued interest of this work.
I am greatly indebted to Professor Alberto Giovannini for interesting
comments and suggestions concerning this work and the positive outlook
for the future.
I would like to thank the organizers of the XXXVII Cracow School of Theoretical
Physics for their invitation and warm hospitality.
% \newpage
%**************
%
%**************
%
%
%

\newpage

\noindent
\underline {{\bf Figure captions}}\\

\noindent
\underline{{\bf Fig.\ 1 }} Feynman diagram for the production  of $m$ gluons
in DLA, where $P$ denotes the 4-momentum of the initial $q (\bar q)$ and
$k_{i}$ denotes the 4-momentum of the ith produced gluon.\\

\noindent
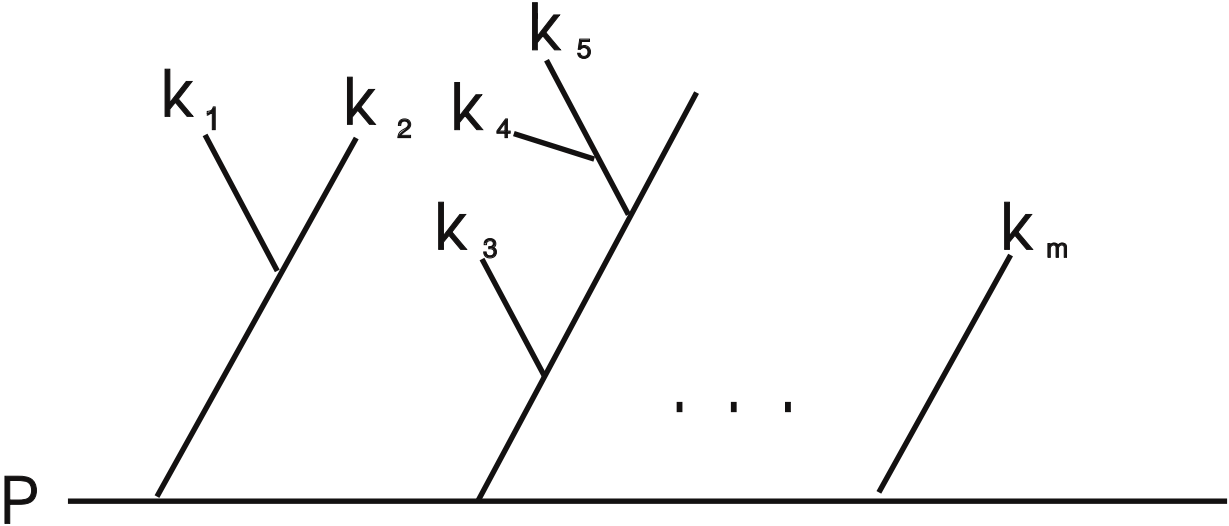
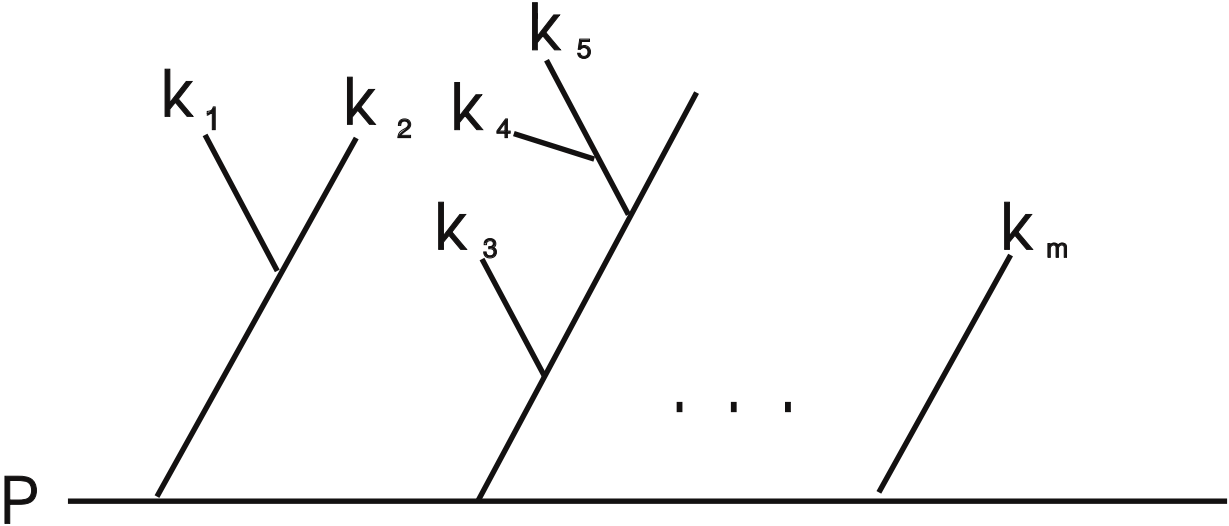
\underline{{\bf Fig.\ 2 }} Generating functional (\ref{16}) as a diagram
series.\\

\noindent
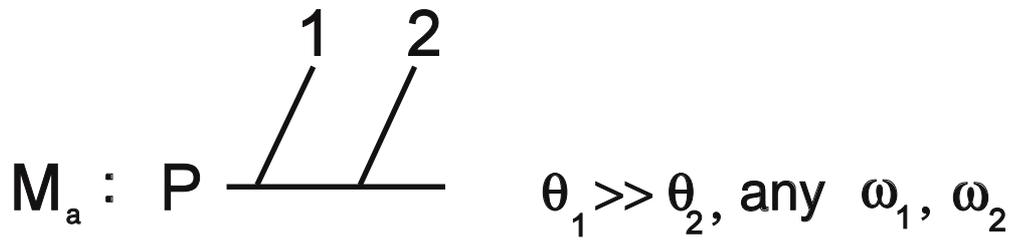
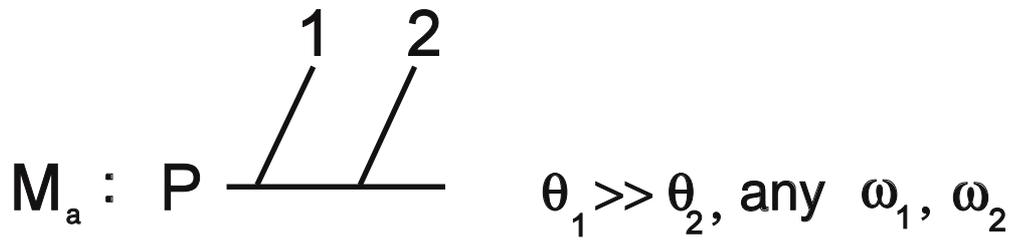
\underline{{\bf Fig.\ 3 }} Redefined kinematical regions in DLA.\\

\noindent
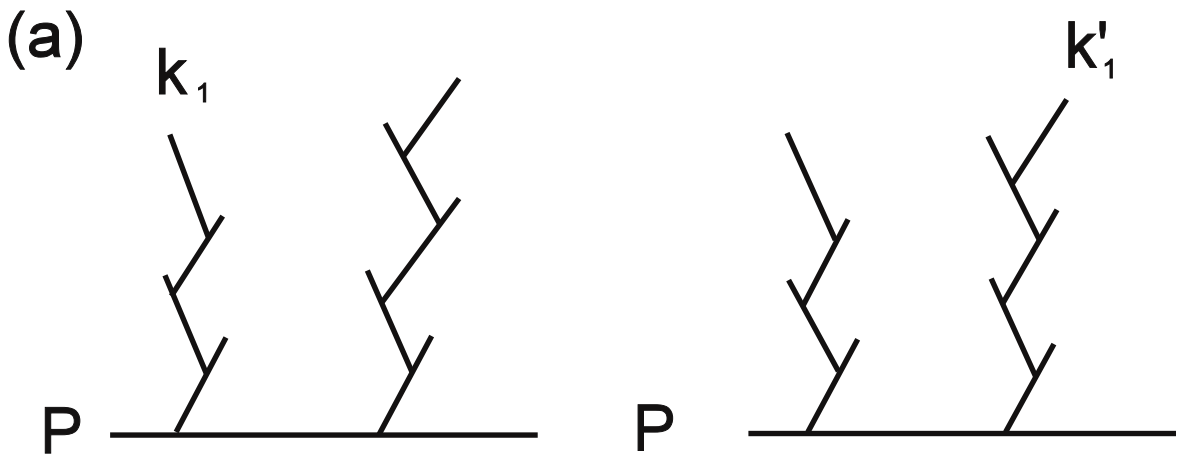
\underline{{\bf Fig.\ 4 }} Interference between different diagrams
for $d^{ex}_{P}(k_{1}',k_{1})$. Remark: diagrams (a) and (b) are identical
except of the position of $k_{1}$ ($k_{1}'$) leg.\\

\noindent
\underline{{\bf Fig.\ 5 }} Master equation for generating functional
(\ref{dla2})
represented as a diagram series (for details see \cite{ziaja}).
Function $P_{1',\ldots,n';1,\ldots,n}$ introduces the parallel angular
ordering (AO).\\

\noindent
\underline{{\bf Fig.\ 6 }} Function $\rho_{P}^{in}({\bf k})$ from
Eq.\ (\ref{p1})
vs $k_{T}\equiv\mid {\bf k}_{T} \mid$ in $Q_{0}$ units
for parameters $b=0.25$, $P/\qu=243$, $\theta=1$, $k/\qu=128$,
chosen following \cite{wos}. The power fit reads $\rho_{P}^{in}({\bf k})=
(\frac{k_{T}}{\qu})^{-1.74}*0.00065\,\qu^{-3}$.
Plot (a) with the logarithmic scale for the
vertical axis, plot (b) with the logarithmic scale for both vertical and
horizontal axes. \\

\noindent
\underline{{\bf Fig.\ 7 }} Function $\rho_{P}^{in,(1)}(k,{\bf x}_{t},0)$ from
Eq.\ (\ref{d4}) vs.\ $x_{T}\equiv\mid {\bf x}_{T} \mid$
for parameters as in Fig.\ 6.
The power fit reads $\rho_{P}^{in}(k,{\bf x}_{T},0)=(\qu x_{T})^{-3.07}*4.0E-05
\,\qu$.
Plot (a) with the logarithmic scale for the
vertical axis, plot (b) with the logarithmic scale for both vertical and
horizontal axes.\\

\noindent
\underline{{\bf Fig.\ 8 }} Function $\rho_{P}^{in}(k,{\bf x}_{t},0)$ from
Eq.\ (\ref{d7}) vs.\ $x_{T}\equiv\mid {\bf x}_{T} \mid$
for parameters as in Fig.\ 6.
Plot (a) with the logarithmic scale for the
vertical axis, plot (b) with the logarithmic scale for both vertical and
horizontal axes.\\
The power fit reads $\rho_{P}^{in}(k,{\bf x}_{T},0)=(\qu x_{T})^{-2.43}*0.00018
\,\qu$.
\\

\noindent
\underline{{\bf Fig.\ 9 }} Diagramatic representation of the term
\bea
& & 2b\,\ln\left(\frac{P}{k}\right)\,
\left\{\int_{0}^{\infty}dx x J_{0}(x\theta_{kk'})\right.
\left.\int_{\frac{\qu}{k}}^{\theta_{Pk}} da J_{0}(xa)\right.
\left.\int_{\frac{\qu}{k}}^{\theta_{Pk'}} da' J_{0}(xa')\right\}=\nonumber\\
& & =\,\int\, d^{3}s \, \frac{k^{3}}{s^{3}}\,A_{ss}(k,{\bf k'}_{T};
k,{\bf k}_{T});
\nonumber
\eea
The integration region lies in the overlap of conus $\Theta_{s}(k)$
and conus $\Theta_{s}(k')$.\\

\noindent
\underline{{\bf Fig.\ 10 }} Diagramatic representation of
$\rho_{P}^{in}(k,{\bf x}_{t},0)$
from Eq.\ (\ref{d7}) as the chain of independent emission factors (\ref{factor})
transformed onto the transverse $x_{T}$ plane by the Bessel factor
of primary emission from the parent P, namely
$J_{0}(\mid {\bf k}_{T} - {\bf k'}_{T}\mid x_{T})$.\\

\begin{thebibliography}{10}
% 0
\bibitem{bz}
B.\ Ziaja, {\sl HEP-PH 9705473}.
% 1
\bibitem{HBT}
R.\ Hanbury-Brown, R.\ Q.\ Twiss, {\sl Nature (London)177, 27 (1956) }.
% 2
\bibitem{gelbke}
D.\ H.\ Boal, C.\ - K.\  Gelbke, B.\ K.\ Jennings, {\sl Rev.\ Mod.\
Phys.\  62, 553 (1990)}.
% 3
\bibitem{others}
W.\ Ochs, J.\ Wosiek, {\sl Phys.\ Letters B305, 144 (1993) }.\\
Y.\ U.\ Dokshitzer, I.\ Dremin, {\sl Nucl.\ Phys.\ B402, 139 (1993)}.\\
P.\ Brax, J.\ -L.\ Meunier, R.\ Peschanski, {\sl Z.\ Phys.\ C62, 649 (1994)}.\\
R.\ Peschanski, {\sl Multiparticle Dynamics 1992, World Scientific 1993}.
% 4
\bibitem{doksh}
Yu.\ L.\ Dokshitzer, V.\ A.\ Khoze, A.\ H.\ Mueller, S.\ I.\ Troyan,
{\sl Basics of Perturbative QCD}, Editions Frontiers, Gif-sur-Yvette Cedex,
France
1991 and references therein.
% 5
\bibitem{lphd}
Ya.\ I.\ Azimov, Yu.\ L.\ Dokshitzer, V.\ A.\ Khoze, S.\ I.\ Troyan, {\sl
Z.\ Phys.\ C27, 65 (1985); C31, 231 (1986)}.
% 6
\bibitem{doksh1}
Yu.\ L.\ Dokshitzer, {\sl Phys.\ Letters B305 (1993), 295} and references
therein.
% 7
\bibitem{pesch1}
J.\ -L.\ Meunier, R.\ Peschanski, { INLN-96-01} and references therein.
% 8
\bibitem{bk}
A.\ Bialas, M.\ Krzywicki, {\sl Phys.\ Letters B354 (1995) 134}.
% 9
\bibitem{ziaja}
B.\ Ziaja, {\sl Acta Phys.\ Polonica B27(1996), 2179}.
% 10
\bibitem{dla}
V.\ S.\ Fadin, {\sl Yad.\ Fiz.\ 37, 408(1983)}.\\
Yu.\ L.\ Dokshitzer, V.\ S.\ Fadin, V.\ A.\ Khoze, {\sl Z.\ Phys.\ C15, 325
(1982), C18, 37 (1983)}.
% 11
\bibitem{wos}
J.\ Wosiek, {\sl Acta Phys.\ Polonica B, 24 (1993), 1027}.\\
W.\ Ochs, J.\ Wosiek, {\sl Phys.\ Letters B305, 144 (1993) }.
% 12
\bibitem{morse}
P.\ M.\ Morse, H.\ Feshbach, {\sl Methods of Theoretical Physics I}, edited
by McGraw-Hill Book Company, Inc.\ 1953, 453
% 13
\bibitem{mueller}
A.\ H.\ Mueller, {\sl Nucl.\ Phys.\ B415 (1994), 373-385 }.\\
% 14
\bibitem{jackson}
J.\ D.\ Jackson, {\sl Classical Electrodynamics}, edited by J. Wiley \& Sons
1975.
\end{thebibliography}
\end{document}